\documentclass[12pt]{article}
\usepackage{latexsym}
\usepackage{amsmath}
\usepackage{multibox}
\usepackage{verbatim}
\usepackage{mathrsfs}
\usepackage{epsfig}
\usepackage{subfigure}

\usepackage{amssymb,amsmath}
\usepackage{amsthm}
\usepackage{amscd}
\usepackage{amssymb}
\usepackage{amsfonts}
\usepackage{url}
\usepackage{slashed}
\usepackage[latin1]{inputenc}
\usepackage[english]{babel}

 \csname
@addtoreset\endcsname{equation}{section}

\def\prd{\pr \cdot}

\def\gz0{\gamma^{0}}

\def\ket#1{|#1\rangle}
\def\scs#1{\section{\sc #1}}
\def\scss#1{\subsection{\sc #1}}

\def\a{\alpha}
\def\b{\beta}
\def\g{\gamma}

\def\d{\delta}

\def\h{\eta}

\def\L{\Lambda}
\def\m{\mu}

\def\s{\sigma}

\def\vf{\varphi}

\def\cA{{\cal A}}
\def\cB{{\cal B}}
\def\cC{{\cal C}}

\def\cF{{\cal F}}
\def\cG{{\cal G}}

\def\cV{{\cal V}}

\def\be{\begin{equation}}
\def\ee{\end{equation}}
\def\bs{\begin{split}}
\def\es{\end{split}}
\def\bea{\begin{eqnarray}}
\def\eea{\end{eqnarray}}
\def\ba{\begin{array}}
\def\ea{\end{array}}
\def\bec{\begin{center}}
\def\ec{\end{center}}
\def\ba{\begin{align}}
\def\ena{\end{align}}

\def\pe{\prime}
\def\12{\frac{1}{2}}
\def\fr{\frac}
\def\pr{\partial}
\def\prd{\partial \cdot}
\def\bra{\langle \,}
\def\ket{\, \rangle}

\def\dsl{\not {\! \pr}}

\begin{document}

\begin{flushright}
{\today}
\end{flushright}

\vspace{25pt}

\begin{center}
{\Large\sc Notes on Strings and Higher Spins}\\ 
\vspace{25pt}
{\sc A.~Sagnotti}\\[15pt]
{\sl\small
Scuola Normale Superiore and INFN\\
Piazza dei Cavalieri, 7\\I-56126 Pisa \ ITALY \\
e-mail: {\small \it sagnotti@sns.it}}\vspace{10pt}
\vspace{35pt}

{\sc\large Abstract}
\end{center}

\noindent These notes are devoted to the intriguing and still largely unexplored links between String Theory and Higher Spins, the types of excitations that lie behind its most cherished properties. A closer look at higher--spin fields provides some further clues that String Theory describes a broken phase of a Higher--Spin Gauge Theory. Conversely, string amplitudes contain a wealth of information on higher--spin interactions that can clarify long--standing issues related to their infrared behavior.

\vskip 3 truecm


\noindent {\sl Based on the lectures presented at the International School for Subnuclear Physics ``Searching for the Unexpected at LHC and Status of Our Knowledge'' (Erice, June 24--July 3 \ 2011), and on the talks presented at ``Strings, Branes and Supergravity'' (Istanbul, July 31--Aug 5 \ 2011), at ``Quantum Theory and Symmetries (QTS7)'' (Prague, Aug. 7--13 \ 2011) and at ``Fundamental Fields and Particles (FFP12)'' (Udine, Nov 21--23 \ 2011). To be reprinted in a special J.~Phys.~A issue devoted to Higher--Spin Theory, eds. M. Gaberdiel and M.A. Vasiliev.}


\setcounter{page}{1}
\pagebreak
{\linespread{0.75}\tableofcontents}
\newpage



\scs{Introduction}\label{sec:Introduction}

More than forty years ago, Veneziano \cite{veneziano} gave birth to String Theory \cite{stringtheory} relating Euler's Beta--function $B(u,v)$ to tree--level S--matrix amplitudes. Together with remarkable subsequent developments, this linked its two arguments $u$ and $v$ to $\a^\prime$, a parameter reflecting a string tension, and to the Mandelstam variables $s$ and $t$ of four--particle amplitudes, in what eventually became tachyon amplitudes of the 26--dimensional bosonic string. This model, however, also contains infinitely many massive excitations, and among them infinitely many higher--spin (HS) ones \cite{solvay} that correspond to multi--symmetric tensors of the type $\vf_{\mu_1 \ldots \, \mu_{s_1}; \, \nu_1 \ldots \, \nu_{s_2};\, \ldots}$. These massive HS excitations are crucial in order to guarantee ``planar duality'', the manifest symmetry of $B(u,v)$ under the interchange of its two arguments, despite their different origins in diagrammatic constructions. For instance, the amplitude built exhibiting \emph{only} $s$--channel poles also possesses $t$--channel poles that can be recovered by resumming polynomial residues, and clearly their infinite number is instrumental to this end.

One can argue, in similar naive terms, that other key properties of String Theory, such as modular invariance or open--closed duality, rest on the presence of infinitely many HS modes. Yet, HS amplitudes have not attracted much attention over the years, possibly because they appear somewhat remote from applications, but certainly because they are rather unwieldy in general. Perhaps this has some bearing on the present, not fully satisfactory, state of String Theory. No neat geometrical principles underlying it have emerged over the years, and this incomplete grasp of the foundations goes along with a lack of universal agreement on the actual lessons that it has in store. Could it be, then, that the massive excitations hold the key for penetrating more deeply into the whole setup? And if this were the case, could the key for understanding the massive excitations lie in the breaking of HS symmetries?
This picture, in fact, has long been in the minds of many authors, at least since the 1980's, when String Field Theory \cite{sftheory} displayed Stueckelberg--like kinetic terms for massive string modes \footnote{The arguments of \cite{bbms}, inspired by the AdS/CFT correspondence \cite{adscft}, are a notable example of this body of evidence.}. Stueckelberg symmetries are indeed deformations of ordinary massless gauge symmetries induced by ``debris'' that is brought about by interactions in the presence of symmetry breaking. As we shall see in the final part of these notes, a closer look at string amplitudes provides further evidence to this effect, although the order parameter, related to $\a^\prime$, lacks a dynamical origin in the current formulation of String Theory.

In mentioning HS fields, I have stressed that they are very complicated in general. As we shall see in the next Section, they predated String Theory by a few decades, since their history started in the 1930's
\cite{majorana,dfp}. Yet, they have long been lagging somewhat behind, so much so that basic ingredients like their free Lagrangian formulations could still receive new inputs during the last decade, while both the nature of their interactions and their actual meaning, let alone their subtle behavior when infra--red cutoffs (masses and/or a cosmological constant) are removed, remain largely mysterious. On the other hand, this framework can exhibit a signpost, the Vasiliev system \cite{vasiliev}, which sits in some respects slightly ahead of those laid by String Theory. Lack of space forces me to leave it out of these notes, so that I can only mention that it describes the mutual interactions of infinitely many symmetric tensors in the presence of a parameter that cuts them off in the infra--red. The actual meaning of this infra--red cutoff is quite interesting in its own right, as stressed in \cite{fms}. The Vasiliev system has the looks of an effective description for the first Regge trajectory of the open bosonic string, in a regime where it should dominate the low--energy dynamics and in a language that extends the frame--like formulation of gravity.

The two main sections that follow are devoted almost exclusively to free and interacting massless HS fields around flat space, the starting point for building massive counterparts in the presence of symmetry breaking. They focus on an extension of the metric--like formulation of gravity that allows a direct comparison with String Theory and begin with some cursory historical remarks that are tailored to the ensuing discussions. Sections \ref{sec:freesymmetric} and \ref{sec:triplets} discuss free irreducible HS fields and their reducible counterparts that emerge from String Theory in the $\a^{\, \prime} \to \infty$ limit, while Section \ref{sec:adsdeformations} discusses free fields in (A)dS backgrounds, stressing role and meaning of their mass--like terms. Section \ref{sec:extcurrents} discusses external currents and the van Dam--Veltman--Zakharov discontinuity, while Section \ref{sec:cubicstringlessons} illustrates some definite lessons that String Theory already provides for HS interactions. Section \ref{sec:beyondcubic}, strictly speaking, would not belong to the Erice course, but it was part of the subsequent talks and I took the freedom to include it since it contains, in my opinion, very interesting clues on the behavior of HS interactions around flat space.


\scs{Free Higher Spins}\label{sec:free}

This Section is devoted to some properties of free symmetric HS tensors, fields of the form $\varphi_{\mu_1 \ldots \, \mu_s}$. Lack of space forces me to confine my attention to Bose fields and to leave out altogether HS fields of mixed symmetry \cite{old,labastida,nlocmixed,mixed}, although they constitute the vast majority of string excitations. The material is meant to prepare the grounds for the discussion of string interactions of Section 3, so that I review Fronsdal's constrained formulation of irreducible HS fields \cite{ff} and its minimal unconstrained extension \cite{fs1}, their links with the geometry of free HS fields and the reducible ``triplets'' \cite{triplet} that emerge from String Theory in the $\a^\prime \to \infty$ limit.

\scss{Free Symmetric Higher Spins}\label{sec:freesymmetric}

It is commonly stated that the theory of HS fields originates from the works of Dirac, Fierz and Pauli \cite{dfp}. Their key results were the elegant DFP physical state conditions
\be
\begin{split}
\left(\Box \, +\, m^2 \right) \varphi_{\mu_1 \ldots \ \mu_s} &= 0 \ , \\
\partial^{\, \mu_1} \, \varphi_{\mu_1 \mu_2 \ldots \ \mu_s} &= 0 \ , \label{DFP1} \\
\eta^{\, \mu_1\mu_2}\, \varphi_{\mu_1 \mu_2 \ldots \ \mu_s} &= 0
\end{split}
\ee
for symmetric tensors $\varphi_{\mu_1 \ldots \ \mu_s}$ of arbitrary rank and their counterparts for symmetric tensor--spinors $\psi_{\mu_1 \ldots \ \mu_s}$ of arbitrary rank
\be
\begin{split}
\left(\dsl \, - \, m \right) \psi_{\mu_1 \ldots \ \mu_s} &= 0 \ , \\
\partial^{\, \mu_1} \, \psi_{\mu_1 \mu_2 \ldots \ \mu_s} &= 0 \ , \label{DFP2} \\
\gamma^{\, \mu_1}\, \psi_{\mu_1 \ldots \ \mu_s} &= 0 \ ,
\end{split}
\ee
whose first members are the Klein--Gordon and Dirac equations, and the first evidence that a naive coupling to Electromagnetism faces unexpected problems. For any given $s$, the first member of the sets \eqref{DFP1} and \eqref{DFP2} defines the mass--shell, the second eliminates unwanted ``time'' components, and finally the last confines the available excitations to irreducible multiplets. Equations of this type lie at the heart of Quantum Field Theory and are key building blocks for most current and past constructions. The logic behind their prominence surfaced long ago in Wigner's work \cite{wigner}: eqs.~\eqref{DFP1} and \eqref{DFP2}, while based on \emph{finite--dimensional non--unitary} representations of the Lorentz group, lead upon quantization to one--particle states filling \emph{unitary} irreducible representations of the Poincar\'e group.

The emergence of higher spins, however, dates back to an older and long neglected paper of Majorana \cite{majorana}. Although this work lies somewhat outside the scope of these notes, I would like to comment briefly on its striking content, since after all the Erice School is rightfully named after this great Italian scientist. Majorana tried to formulate a Dirac--like equation free of the then disturbing negative--energy solutions, and in doing so was led to consider \emph{infinite--dimensional unitary} representations of the Lorentz group SO(3,1). Let me stress that in 1932 this subject was largely unfamiliar in Mathematics, let alone in Theoretical Physics. Majorana readily found out that, in the massive case, his equation was not describing a single type of particle, but rather a whole ``trajectory'' of states with
\be
M(s)\ \sim \ \frac{1}{s + \frac{1}{2}} \ .
\ee
He was thus anticipating, to some extent, both Regge's idea \cite{regge} and its eventual realization in the Veneziano amplitude \cite{veneziano}, and thus in String Theory altogether \cite{stringtheory}, by over thirty years! On the mathematical side, he had come across an oscillator realization of the $so(3,1)$ algebra, well before Mathematics touched upon a tool that only in the 1970's became relatively familiar in Theoretical Physics \cite{gunaydin}. The paper is striking, since its techniques are a sublimation of elementary facts of angular momentum theory that were then common tools in Atomic Physics. His analysis rests on oscillator generalizations of the Dirac $\gamma$--matrices, but it was perhaps too early even for him to appreciate that he was actually building, at no extra cost, unitary representations of the larger $so(3,2)$ algebra. The extension reflects, in modern terms, a familiar fact: if one combines $\gamma^\mu$'s and $\gamma^{\mu\nu}$'s, the resulting commutators generate indeed $so(3,2)$. In fact, without being fully aware Majorana had stumbled upon Dirac's ``singletons'' \cite{diracsing}, which lie at the heart of beautiful sequels by Flato, Fronsdal and others \cite{flatofronsdal}, and more recently of both the AdS/CFT correspondence \cite{adscft} and Vasiliev's theory \cite{vasiliev}. I will end here these parenthetic remarks, referring the interested reader to \cite{majorana} and to some interesting recent papers \cite{casalbuoni} on Majorana's 1932 contribution, where the subsequent history of the subject is also addressed.

The discovery of the Rarita--Schwinger theory \cite{RS}, a development that played a key role in Theoretical Physics toward the end of the last Century as a key building block of supergravity \cite{supergravity}, followed rather closely the Schr\"odinger--like DFP conditions \eqref{DFP1} and \eqref{DFP2}, but World War II forced the scientific community to leave aside these themes for a while. As a result, a systematic search for action principles, the next natural task along this line, had to await about four decades, since in the 1950's and 1960's, when a sizable activity on fundamental questions resumed, the focus was rather on the $S$--matrix, with an eye on important issues in current algebra.

Rather than spurring progress in a constructive sense, key developments on HS that took place in the 1960's actually spotted further unexpected difficulties. In retrospect, these results were unveiling important information on the nature of HS interactions, but for a while their net effect was to force most of the community away from these themes. I am referring, in particular, to Weinberg's 1964 argument on soft HS particles and the $S$ matrix \cite{weinb64}, to the Coleman--Mandula theorem \cite{cm} and to the Velo--Zwanziger violation of causality in the presence of electromagnetic interactions \cite{vz}. These were then followed, after a few years, by the Aragone--Deser field--theory argument \cite{ad} on the conflict between ``minimal'', two--derivative, gravitational couplings and HS fields around flat space, and by the Weinberg--Witten theorem \cite{ww,porratiww}, a result that points in the same direction but is again inspired by $S$--matrix techniques. I will return to these issues in Section \ref{sec:interacting}.

Some major developments that took place in Field Theory during the 1970's were certainly not foreign to the renewed interest that ultimately led to action principles for massive symmetric Bose and Fermi HS fields \cite{SH}. These results were the starting point for the subsequent work by Fronsdal and Fang and Fronsdal \cite{ff}, where gauge--invariant actions were finally recovered in the massless limit.

Fronsdal arrived at a natural generalization of the Maxwell and linearized Einstein equations. He extended the familiar $s=1,2$ cases
\be
\begin{split}
& \Box A_\mu \, -\, \partial_\mu \, \partial \cdot A \, = \, 0 \ , \\
& \Box \varphi_{\mu\nu} \, - \, \left( \, \partial_\mu \, \partial \cdot \varphi_\nu \, + \, \partial_\nu \, \partial \cdot \varphi_\mu \, \right) \, + \,
\partial_\mu \, \partial_\nu \, \varphi^{\; \prime} \, = \, 0 \ , \label{s12}
\end{split}
\ee
where the ``prime'' denotes a trace, that are invariant under the gauge transformations
\be
\begin{split}
& \delta A_\mu \, =\, \partial_\mu \, \Lambda \ , \\
& \delta \, \varphi_{\, \mu\nu} \, = \, \partial_\mu \, \Lambda_\nu \, + \, \partial_\nu \, \Lambda_\mu \ , \label{gauge12}
\end{split}
\ee
to symmetric tensors $\varphi_{\mu_1 \ldots \, \mu_s}$ of arbitrary rank, arriving at the second--order equations
\be
{\cal F}_{\, \mu_1 \ldots \, \mu_s\,}\, \equiv \, \Box \, \varphi_{\, \mu_1 \ldots \, \mu_s} \, - \, \left( \, \partial_{\mu_1} \, \partial \cdot \varphi_{\, \mu_2 \ldots \, \mu_s} \, +\, \ldots \, \right) \, + \,
\left( \, \partial_{\mu_1} \, \partial_{\mu_2} \, \varphi^{\; \prime}_{\, \mu_3 \ldots \, \mu_s} \, +\, \ldots \, \right)\, = \, 0 \ , \label{ss}
\ee
where the combinations ${\cal F}_{\, \mu_1 \ldots \, \mu_s\,}$ are often referred to as Fronsdal's tensors.
As these are natural generalizations of eqs.~\eqref{s12}, one would expect that eqs.~\eqref{gauge12} leave way to the natural spin--$s$ gauge transformations
\be
\delta \varphi_{\, \mu_1 \ldots \, \mu_s} \, = \, \partial_{\mu_1} \, \Lambda_{\, \mu_2 \ldots \, \mu_s} \, +\, \ldots \ , \label{gaugess}
\ee
but as we shall see shortly this is not quite the case.

In eqs.~\eqref{ss} and \eqref{gaugess}, a number of terms needed to complete the symmetrizations are inevitably left implicit. In order to streamline the ensuing discussion, it is very convenient to resort to a notation introduced in \cite{fs1} that may be regarded as a counterpart, for these symmetric tensors, of the language of forms. Lorentz labels are thus eliminated altogether, so that a generic spin--$s$ field is simply denoted by $\varphi$, and in this fashion eqs.~\eqref{ss} and \eqref{gaugess} take the simpler forms
\be
{\cal F} \, \equiv \, \Box \, \varphi \, - \, \partial \, \partial \cdot \varphi \, + \,
\partial^{\, 2} \, \varphi^{\; \prime} \, = \, 0 \ , \label{ssred}
\ee
and
\be
\delta \varphi \, = \, \partial \, \Lambda \ . \label{gaugessred}
\ee

Let me stress that this choice is not merely a shorthand. Not only does it simplify greatly all expressions at hand, in fact, but it also embodies the key properties that are needed to manipulate them at will. To this end, one only needs a handful of rules, that are discussed in some detail, for instance, in \cite{fms}:
\be
\begin{split}
  \left( \pr^{\, p} \, \vf  \right)^{\, \pe} & = \ \Box \,
  \pr^{\, p-2} \, \vf \ + \, 2 \, \pr^{\, p-1} \,  \prd \vf \ + \, \pr^{\, p} \,
\vf^{\, \pe} \,  ,  \\
 \partial^{\, p} \, \partial^{\, q} & = \ \binom{p+q}{p} \ \
\partial^{\, p+q} \ , \\
 \partial \cdot  \left( \partial^{\, p} \ \vf \right) & = \ \Box \
\partial^{\, p-1} \ \vf \ + \
\partial^{\, p} \ \partial \cdot \vf \ ,  \\
 \partial \cdot  \left(\eta^{\, k}\vf \right) & = \ \partial \, \eta^{\, k-1}\vf \ + \ \eta^{\, k} \, \left(\partial \, \cdot \vf \right) \ , \\
 \left( \eta^k \, \vf  \,  \right)^{\, \prime} & = \ \left[ \, D
\, + \, 2\, (s+k-1) \,  \right]\, \eta^{\, k-1} \, \vf \ + \ \eta^k
\, \vf^{\, \prime} \,  . \label{etak}
\end{split}
\ee

Suffice it to say, here, that in this notation every implicit symmetrization is meant to involve the \emph{least} number of terms that are needed to effect it, and that the overall coefficients are always one. One can thus understand, for instance, why $\partial \, \partial = 2\, \partial^{\, 2}$, the counterpart of the vanishing of the squared exterior derivative: since derivatives commute, a naive symmetrization would overcount the terms actually needed by a factor two.

Returning to the Fronsdal equations \eqref{ss}, their structure is clearly quite natural in view of eqs.~\eqref{s12}, and yet their behavior under gauge transformations entails a little surprise, since
\be
\delta \, {\cal F} \, = \, 3 \, \partial^{\, 3} \Lambda^{\, \prime} \ . \label{gauge_unconstr}
\ee
In other words, gauge invariance demands that $\Lambda^{\, \prime} = 0$, \emph{i.e.} that the parameters be \emph{traceless}.
This condition first shows up for $s=3$ and is often referred to as \emph{Fronsdal's first constraint}. A second, subtler constraint, is also needed in order that the natural counterpart of the free Einstein Lagrangian, that up to partial integrations can be presented in the form
\be
{\cal L} \, = \, \varphi \left( {\cal F} \, - \, \frac{1}{2} \ \eta \, {\cal F}^{\, \prime} \right) \ ,
\ee
be gauge invariant, even with traceless parameters. As we have seen, in this case ${\cal F}$ is gauge invariant, while up to total derivatives the variation of the ``naked'' $\varphi$ yields
\be
\delta{\,\cal L} \, = \,  - \, s \, \Lambda \left( \partial \cdot {\cal F} \, - \, \frac{1}{2} \ \partial {\cal F}^{\, \prime} \right) \ , \label{lags}
\ee
an expression that would vanish identically if a HS counterpart of the familiar Noether or Bianchi identity for the linearized version of Einstein's gravity,
\be
\partial^{\,\mu} \, {\cal R}_{\mu\nu} \, - \, \frac{1}{2} \ \partial_\nu \, {\cal R} \, = \, 0 \ ,
\ee
were to hold. As it turns out, however, this is not the case, since in general
\be
\partial \cdot {\cal F} \, - \, \frac{1}{2} \ \partial \, {\cal F}^{\, \prime} \, = \, - \, \frac{3}{2} \ \partial^{\, 3} \, \varphi^{\, \prime\prime} \, ,
\ee
so that for $s\geq4$ gauge invariant action principles obtain only if the gauge fields are subject to \emph{Fronsdal's second constraint}, $\varphi^{\, \prime\prime} = 0$, the condition that $\varphi$ be \emph{doubly traceless}.

The need for these constraints is a sign that Fronsdal's formulation is incomplete, since eqs.~\eqref{ss} and \eqref{lags} do not accommodate the natural gauge symmetry \eqref{gaugess}, despite the fact that algebraic constraints certainly do not raise doubts on the nature of the actual propagating degrees of freedom. This problem was first bypassed in the 1990's by the Dubna group \cite{dubna}, adapting to free higher spins the BRST construction \cite{brst} of free String Field Theory \cite{sftheory}, but at the price of introducing ${\cal O}(s)$ different fields for the case of a rank--$s$ tensor $\varphi_{\mu_1 \ldots \mu_s\,}$. A ``minimal'' solution was then obtained in \cite{fs1,fms}. Its ``compensator equations''
\be
{\cal A} \, \equiv \, {\cal F} \, - \ 3 \, \partial^{\, 3} \alpha \, = \, 0 \label{ssredc}
\ee
involve, for $s \geq 3$, only one additional field, a spin--$(s-3)$ compensator $\alpha_{\mu_1 \ldots \, \mu_{s-3}}$, and the resulting extension ${\cal A}$ of Fronsdal's ${\cal F}$ tensor is gauge invariant provided
\be
\delta \alpha \, = \, \Lambda^{\, \prime} \ . \label{deltacomp}
\ee

Actually, for $s=3$ the compensator made an early appearance in the work of Schwinger \cite{schwinger}, as I was told some time ago by G.~Savvidy. Schwinger's books contain indeed a discussion, for $s=3$, of this additional field, which he also introduced in order to bypass Fronsdal's first constraint but readily abandoned since it brings along higher derivatives. In retrospect, these higher derivatives are harmless since they affect non--propagating components, and we now understand how to trade them for a few more fields, whose number does not grow with the spin, as in \cite{lowder,bgk}.

An additional field, a Lagrange multiplier $\beta_{\mu_1 \ldots \, \mu_{s-4}}$ such that
\be
\delta \, \beta \, = \, \partial \cdot \partial \cdot \partial \cdot \Lambda \ ,
\ee
is also present for $s \geq 4$ in the complete Lagrangians \cite{fs1,fms}
\be
{\cal L} \, =\, \varphi\left(  {\cal A} \, - \, \frac{1}{2} \ \eta \, {\cal A}^\prime \right)
\, - \, \frac{3}{4} \, \binom{s}{3} \ \alpha \, \partial \cdot
{\cal A}^\prime \, + \, 3 \, \binom{s}{4} \, \beta \, {\cal C} \ ,
\ee
where
\be
\cC \, = \, \varphi^{\, \prime\prime} \, - \, 4 \, \partial \cdot \alpha \, - \, \partial \, \alpha^{\, \prime}
\ee
is a gauge--invariant completion of the double trace of $\varphi$. The resulting Lagrangian equations for $\vf$ are
\be \label{boseeom}
\begin{split}
& \cA\, - \, \12 \, \h\, \cA^{\, \pe} \,
\, + \, \h^{\, 2} \, \cB \, = \, 0 \, ,    \\
& \varphi^{\, \prime\prime} \, - \, 4 \, \partial \cdot \alpha \, - \, \partial \, \alpha^{\, \prime} \, = \, 0 \, ,
\end{split}
\ee
where $\eta$ is the Minkowski metric and
\be \cB \, = \, \b \, - \, \12 \, (\prd \prd \vf^{\, \pe} \, -
\, 2 \,  \Box \, \prd \a \, - \, \pr \, \prd \prd \a)  \ee
is a gauge invariant completion of the Lagrange multiplier $\beta$. Let me stress that eqs.~\eqref{boseeom} generalize the linearized Lagrangian equation for gravity,
\be
{\cal R}_{\mu\nu} \, - \, \frac{1}{2} \ \eta_{\mu\nu} \ {\cal R} \, = \, 0 \ ,
\ee
and are thus more complicated than eq.~\eqref{ssred} that is rather the counterpart for these systems of the simpler linearized non--Lagrangian equation
\be
{\cal R}_{\mu\nu} \, = \, 0 \ .
\ee
A recursive argument \cite{fs1,fms} shows that eqs.~\eqref{boseeom} can turned into
\be
\cB \, = \, 0 \qquad {\rm and} \qquad
\cA \, = \, 0 \label{nonlags}
\ee
once they are combined with their traces.

Removing Fronsdal's constraints via the compensator $\alpha$ (and the Lagrange multiplier $\beta$) is clearly eliminating an asymmetry between the first two cases, $s=1,2$, and the HS ones. Interestingly, this also opens a small and yet instructive window on the geometrical nature of HS constructions. After all, eqs.~\eqref{s12} admit a different but equally familiar presentation in terms of the Maxwell curvature $F_{\mu\nu}$ and, as we have already anticipated, of its counterpart ${\cal R}_{\mu\nu}$ for the linearized Einstein theory. There is a key difference between these objects, however: for $s=2$ one arrives at the linearized Riemann curvature tensor ${\cal R}_{\mu\nu\rho\sigma}$ via an intermediate object, the Christoffel connection, that in the linearized case reads
\be
{\Gamma^\mu}_{\nu\rho} \, = \, \frac{1}{2} \left( \partial_\nu \, {\varphi^\mu}_\rho \, + \, \partial_\rho \, {\varphi^\mu}_\nu \, - \,
\partial^\mu \, \varphi_{\nu\rho}\right) \ .
\ee
As a result, while the Maxwell curvature contains one derivative of $A_\mu$, its $s=2$ counterpart is bound to contain two derivatives of $\varphi_{\mu\nu}$.

It is convenient to focus here on a variant of the usual Riemann curvature that is \emph{symmetric} under interchanges within its two groups of indices, although it is merely a combination of conventional Riemann tensors on account of the cyclic identity. Yet, when proceeding to arbitrary values of $s$, this choice has the virtue of suggesting a recursive construction where
the linearized $\Gamma_{\mu, \, \nu\rho}$ leaves way to a tower of $s-1$ Christoffel--like connections, $\Gamma_{\mu_1 , \, \nu_1 \ldots \, \nu_s}$,
\ldots ,
$\Gamma_{\mu_1 \ldots \, \mu_{s-1} , \, \nu_1 \ldots \, \nu_s}$. Each of these quantities is obtained from combinations of derivatives of the previous member of the list in such a way that, in the gauge transformations, all $\mu$ indices fall on the gauge parameter. These expressions carry, by construction, 1, \ldots , $s$-1 derivatives of the original field, and the next member of the sequence is a spin--$s$ analogue of the linearized $s=2$ curvature in its doubly--symmetric incarnation, a tensor
${\cal R}_{\mu_1 \ldots \, \mu_s , \, \nu_1 \ldots \, \nu_s}$ built from combinations of $s$ derivatives of the original $\varphi_{\mu_1 \ldots \mu_s\,}$  that is \emph{gauge invariant} under transformations like \eqref{gaugess}, albeit with unconstrained parameters. One can also show that these curvature tensors possess the additional symmetry
\be
{\cal R}_{\mu_1 \ldots \, \mu_s , \, \nu_1 \ldots \, \nu_s} = (-1)^{\, s} \ {\cal R}_{\nu_1 \ldots \, \nu_{s} , \, \mu_1 \ldots \, \mu_s} \ ,
\ee
which is already visible in the $s=1,2$ cases.

The beautiful construction that I just sketched, due to de Wit and Freedman \cite{dwf}, marks the real entry point of HS geometry into the game. Since, as I have stressed, the relevant HS curvatures are by construction invariant under unconstrained gauge transformations of the type \eqref{gaugess}, in this fashion the authors came short of bypassing Fronsdal's constraints. They did not make this further step, however, since they insisted on the apparently natural condition that Fermi or Bose equations contain, respectively, one or two derivatives. As a result, while they could recognize that the Fronsdal tensor ${\cal F}$ is related to the second Christoffel--like connection according to
\be
{\cal F}_{\nu_1 \ldots \, \nu_s} \, \sim \, \eta^{\mu_1 \mu_2} \ \Gamma_{\mu_1 \mu_2 , \, \nu_1 \ldots \, \nu_s} \ ,
\ee
all in all their HS geometry appeared for a while somewhat remote from the actual free dynamics.

The compensator equations do better in this respect, as they should since they embody an unconstrained gauge symmetry. Francia and I obtained these equations directly \cite{fs1}, before formulating the compensator theory, but it is instructive to retrace the argument in this fashion, starting from the compensator equations. Let us do it in some detail for the simplest case, $s=3$, in which the non--Lagrangian equations \eqref{nonlags} reduce to
\be
{\cal F}_{\mu\nu\rho} \, = \, 3 \, \partial_\mu \partial_\nu \partial_\rho \, \alpha \ .
\ee
One can formally solve this equation for $\partial_\mu \, \alpha$, obtaining
\be
\partial_\mu \, \alpha \, = \, \frac{1}{3} \ \frac{1}{\Box} \ {{\cal F}^{\, \prime}}_{\mu} \ , \label{nloc1}
\ee
or directly for $\alpha$, obtaining
\be
\alpha \, = \, \frac{1}{3} \ \frac{1}{\Box^{\, 2}} \ \partial \cdot {{\cal F}^{\, \prime}} \ , \label{nloc2}
\ee
where I have resorted again to the shorthand ``prime'' notation to indicate traces.
This is tantamount to turning the compensator equation into a pair of distinct \emph{non--local} equations for $\varphi_{\mu\nu\rho}$ alone. By construction, the end result is invariant under the unconstrained gauge transformations \eqref{gaugess}, but it is clearly \emph{not} unique, since eq.~\eqref{nloc1} leads to
\be
{{\cal F}^{(1)}}_{\mu\nu\rho} \, \equiv \, {\cal F}_{\mu\nu\rho} \, - \, \frac{1}{3\, \Box} \, \big( \partial_\mu \, \partial_\nu \, {{\cal F}^{\, \prime}}_{\rho} \, + \, \partial_\nu\, \partial_\rho \, {{\cal F}^{\, \prime}}_{\mu} \, +\, \partial_\rho \, \partial_\mu \, {{\cal F}^{\, \prime}}_{\nu}\big) \, = \, 0 \ , \label{s3nl1}
\ee
while eq.~\eqref{nloc2} leads to the alternative gauge invariant non--Lagrangian equation
\be
{\widetilde{\cal F}^{(1)}}_{\mu\nu\rho} \, \equiv \, {\cal F}_{\mu\nu\rho} \, - \, \frac{1}{\Box^2} \ \partial_\mu \, \partial_\nu \, \partial_\rho \, \partial \cdot \, {\cal F}^{\, \prime} \, = \, 0 \ .
\label{s3nl2}
\ee

Having more than one option might seem problematic, but as we shall see it is actually instrumental for the very consistency on the non--local formulation, and to begin with the reader can verify that eqs.~\eqref{s3nl1} and \eqref{s3nl2} can be turned into one another, so that they are nicely equivalent in the absence of external sources. Interestingly, factoring out the inverse d'Alembertian present in eq.~\eqref{s3nl1} according to
\be
{{\cal F}^{(1)}}_{\mu\nu\rho} \, \equiv \, \frac{1}{\Box} \left[ \Box \, {\cal F}_{\mu\nu\rho} \, - \, \frac{1}{3} \, \left( \partial_\mu \, \partial_\nu \, {{\cal F}^{\, \prime}}_{\rho} \, + \, \partial_\nu\, \partial_\rho \, {{\cal F}^{\, \prime}}_{\mu} \, +\, \partial_\rho \, \partial_\mu \, {{\cal F}^{\, \prime}}_{\nu}\right)\right] \, = \, 0 \ , \label{s3nl12}
\ee
leaves room for a total of four derivatives within the square brackets, as many as the divergence of the spin--3 curvature would contain, according to the preceding discussion. All in all, one is thus led to the conclusion that the spin--3 non--local equations of motion, and thus a fortiori their local compensator counterparts, are equivalent to
\be
\frac{1}{\Box} \ \eta^{\, \mu_1\mu_2}\, \partial^{\, \mu_3} \ {\cal R}_{\mu_1\mu_2\mu_3,\nu_1\nu_2\nu_3} \, = \, 0 \ ,
\label{nlocs3}
\ee
with ${\cal R}$ the spin--3 de Wit--Freedman curvature, symmetric in its two sets of Lorentz labels and antisymmetric under an overall interchange of them. This can be simply verified, and clearly eq.~\eqref{nlocs3} is a very natural and satisfactory spin--3 counterpart of the Maxwell equation, albeit \emph{not} a Lagrangian equation.

This pattern extends to arbitrary spin, as can be appreciated via an inductive argument that rests on the sequence of integro--differential operators
\be
{\cal F}^{(n+1)} = {\cal F}^{(n)} + \frac{1}{(n+1)(2n+1)}\ \frac{\partial^2}{\Box} \ {\cal F}^{\prime (n)} -
\frac{1}{n+1} \ \frac{\partial}{\Box} \ \partial \cdot {\cal F}^{(n)} \ , \label{iteration}
\ee
whose Bianchi identities take the form
\be
\prd {\cal F}^{(n)} \ - \ \frac{1}{2n} \ \pr \, {{\cal F}^{(n)}}{\; '} \ = \ - \
\left( 1 + \frac{1}{2n}  \right) \ \frac{\pr^{\; 2n+1}}{\Box^{\; n-1}} \
\vf^{[n+1]} \ , \label{bianchin}
\ee
where $\vf^{[n+1]}$ denotes the $(n+1)$--st trace of the HS field $\vf$.
For these combinations, the original compensator equation \eqref{ssred} translates into the sequence
\be
{\cal F}^{(k)} \ = \ (2 \, k \ + \ 1) \ \frac{\partial^{\, 2
k+1}}{\Box^{k-1}} \ \alpha^{[k-1]}\ , \label{compensk}
\ee
where $\alpha^{[k-1]}$ denotes the $(k-1)$--fold trace of $\a$,
whose members have thus the virtue of involving successive traces of the compensator. For any given $s$, these traces are simply not available for $k$ sufficiently large, so that fully gauge invariant equations obtain after completing $\left[\frac{s}{2}\right]$ iterations. Taking into account for any $s$ the first of these equations, one is finally led to an infinite family of fully gauge invariant \emph{non--local} counterparts of the Maxwell and Einstein equations,
\be
\begin{split}
& s=2n+1 : \quad \frac{1}{\Box^n} \ \partial_\mu {\cal R}^{\mu [n];\nu_1 \ldots \nu_s} \, = \, 0 \ ,  \\
& s = 2n : \quad \quad \ \ \ \frac{1}{\Box^{n-1}}\  {\cal R}^{[n];\nu_1 \ldots \nu_s}\, =\,0 \ . \label{nloceqs}
\end{split}
\ee

While these non--Lagrangian equations involve natural geometric structures for these systems in the simplest combinations that one could a priori conceive, the corresponding Lagrangian equations rest not only on traces of the first gauge--invariant ${\cal F}^{(k)}$ but also on higher members of the set, obtained via further iterations of eq.~\eqref{iteration}. We shall return shortly to this point, after introducing some additional tools.

Before closing this section, let me stress briefly that what I have said extends to Fermi fields carrying a set of totally symmetric vector labels, up to standard
complications brought about by $\gamma$--matrices \cite{ff,dubna,fs1,fms,lowder}. Tensors and spinor--tensors of mixed symmetry, on the other hand, rest on a richer theory that originates from the work of Labastida and others \cite{old,labastida} and would require a detailed discussion. I will content myself with stating that one can extend to this case the non--local equations \eqref{nloceqs} \cite{nlocmixed} and, up to some subtleties that are discussed in detail in \cite{mixed}, the compensator formulation as well. The subject is very interesting and is crucial in order to come eventually to terms with String Theory, but lack of space forces me to leave it out here.

\scss{Triplets and free String Field Theory}\label{sec:triplets}

A key question that these notes are meant to address is what these notions about higher spins are teaching us about the massive string excitations.

At the free level, the link between String Theory and HS is relatively simple, and yet it is interesting and quite instructive. In short, String Theory favors reducible representations, so that it drops somehow the last DFP condition of eq.~\eqref{DFP1}. For symmetric Bose fields this entails the replacement of the Fronsdal Lagrangian or its compensator extensions with an interesting system of three fields that is often called a ``triplet'' \cite{triplet,dubna,fs1,st1,francia10} in the literature. The three fields are a rank--$s$ tensor $\varphi_{\mu_{\,1} \ldots \, \mu_{\,s}}$, the dynamical field in this context, and two additional tensors of ranks $s-1$ and $s-2$, $C_{\mu_{\,1} \ldots \, \mu_{\,s-1}}$ and $D_{\mu_{\,1} \ldots \, \mu_{\,s-2}}$ that disappear on shell but are nonetheless crucial to attain an \emph{unconstrained} gauge symmetry. The $s=1$ case is degenerate, and the system reduces to the description of an electromagnetic potential in the Nakanishi--Lautrup presentation.

In the index--free notation, the triplet equations read
\be
\begin{split}
& \Box \; \varphi \ = \ \partial \, C \ ,  \\
& \partial \cdot \varphi \ - \ \partial \, D \ = \ C  \ , \\
& \Box \; D \ = \ \partial \cdot C \ , \label{triplet}
\end{split}
\ee
and are invariant, as anticipated, under the \emph{unconstrained} gauge transformations
\be
\begin{split}
& \delta \; \varphi \ = \ \partial \, \Lambda \ ,  \\
& \delta \; C \ = \Box \, \Lambda  \ , \\
& \delta \; D \ = \ \partial \cdot \Lambda \ .
\end{split}
\ee
If one is ready to take for granted that $C$ and $D$ disappear on shell, which can be shown in a few steps, the triplet equations \eqref{triplet} clearly boil down to the first two DFP conditions \eqref{DFP1}.

The triplet equations \eqref{triplet} follow from the elegant BRST \cite{brst} formulation of String Field Theory \cite{sftheory}. Its field equation,
\be
{\cal Q} \, | \Psi \rangle \, = \, 0 \ ,
\ee
rests on the BRST operator ${\cal Q}$ of world--sheet reparametrizations, whose nilpotency in the critical dimension brings about the unconstrained gauge symmetry
\be
\delta \, | \Psi \rangle \, = \, {\cal Q}\, | \Lambda \rangle \ ,
\ee
together with gauge--for--gauge counterparts. However, while the string BRST operator is only nilpotent in the critical dimension, its contracted version that emerges in the formal limit $\a^\prime \to \infty$ is \emph{identically} nilpotent \cite{dubna}, and as a result the triplet system \eqref{triplet} is indeed gauge invariant in any number of dimensions.
An alternative presentation of eqs.~\eqref{triplet} obtains if $C$ is eliminated algebraically via the second of them, reducing the first to
\be
{\cal F} \, = \, \partial^{\, 2} \, \left( \varphi^{\, \prime} \, - \, 2\, D \right) \ . \label{tripletred}
\ee
One can simply select the spin--$s$ excitations of this system \cite{fs1}, adding to eqs.~\eqref{triplet} the further condition
\be
\varphi^{\, \prime} - 2\, D \, = \, \partial \, \alpha \  \label{hstriplet}
\ee
that enforces on shell, in a gauge invariant fashion, the last DFP condition of eq.~\eqref{DFP1}. This step brings about a compensator $\alpha$ with the gauge transformation of eq.~\eqref{deltacomp}, while eq.~\eqref{tripletred} turns readily into eq.~\eqref{ssred}. Following \cite{bgk}, one can also perform this reduction at the Lagrangian level, at the price of a few additional fields whose number, however, does not grow with the spin.

Another interesting question concerns the non--local formulation of the triplet systems, based on $\varphi$ alone. One can build it systematically \cite{francia10} and the end results,
\be
{\cal L} \ \sim \ {\cal R}^{ \mu_1 \ldots \, \mu_s \, ; \, \nu_1 \ldots \, \nu_s} \, \frac{1}{\Box^{s-1}} \  {\cal R}_{\mu_1 \ldots \, \mu_s\, ; \, \nu_1 \ldots \, \nu_s} \ , \label{lagnlred}
\ee
are strikingly natural non--local counterparts of the Maxwell Lagrangian.


\scs{Interacting Higher Spins}\label{sec:interacting}

We can now turn to HS interactions, and to begin with I would like to comment on some classic results of the 1960's and 1970's that revealed unexpected difficulties.
\begin{figure}
\begin{center}
\epsfig{file=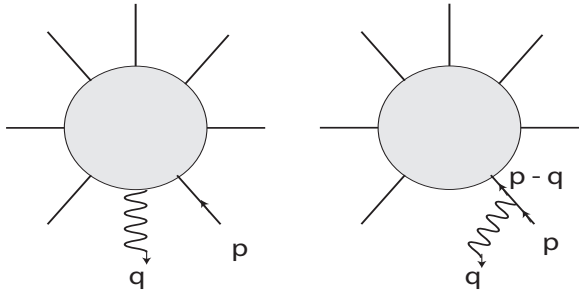, height=1.5in, width=3in}
\caption{Factorization of an amplitude containing a ``soft'' emission.}
\end{center}
\label{soft}
\end{figure}

The first and most striking of these results was obtained by S.~Weinberg in 1964 \cite{weinb64}. It concerns the behavior that S--matrix amplitudes for a single \emph{massless} spin--$s$ particle and a number of scalar particles would exhibit in the presence of long--range HS interactions. Referring to fig.~1, the ``soft'' limit $q \to 0$ of one such amplitude, here denoted by $A_s$, would expose long--range HS effects and would be dominated by poles proportional to $p_i \cdot q$ originating from scalar propagators close to their mass shells. Moreover, Lorentz--invariant vertices for two scalars and a spin--$s$ particle giving rise to long--range HS effects would lead, in the ``soft'' limit, to $s$ powers of $p_i$ contracting the HS polarization tensor. For $q \to 0$, an $N$--point $A_s$ amplitude would thus behave asymptotically as
\be
A_s(1,\ldots\,,N) \ \sim \ A(1,\ldots\,,N-1) \, \times \, \sum_{i=1}^{N-1} \frac{p_{i,\,\mu_1} \ldots p_{i,\,\mu_s} \ \vf^{\,\mu_1\ldots \mu_s}(q)}{2 \,p_i \cdot q} \ , \label{wei64}
\ee
where $A(1,\ldots\,,N-1)$ denotes a more conventional $(N-1)$--point scalar amplitude, the residue of the pole.

Weinberg's key observation was that, although an \emph{on--shell} Abelian gauge transformation
\be
\vf(q) \to \vf(q) \ + \ i\, q\, \Lambda(q)
\ee
of the spin--$s$ polarization tensor would seem to cause to no problems due the ``soft'' nature of the momentum $q$, the pole would make the end result \emph{finite} in the limit. Consequently, HS amplitudes might well be \emph{not} invariant under these shifts, with the disastrous consequence that unphysical polarizations might \emph{not} decouple. Unless, of course, some conditions are met, and a clear way around does exists, but only for $s=1$ or $2$. In the former case, the limiting contribution \eqref{wei64} would be indeed proportional to the total electric charge of the (ingoing -- outgoing) scattered particles, which ought to vanish in $S$--matrix amplitudes, while in the latter it would be proportional to the sum of the scalar momenta weighted by the corresponding ``gravitational charges''. Hence, with gravity coupling universally, for $s=2$ consistency would translate into momentum conservation and therefore would be automatically guaranteed. Conversely, the requirement of consistency for long--range gravitational interactions emerges, in this fashion, as a reason for their universality. On the other hand, for $s \geq 3$ eq.~\eqref{wei64} would involve polynomials in the momenta, that clearly would not add up to zero in general. All in all, generic amplitudes with a single \emph{massless} HS particle would thus seem inconsistent, but there are possible loopholes. First, as noticed in \cite{weinb64}, if more powers of q were present in the vertices, they could compensate the pole, making the terms in eq.~\eqref{wei64} vanish individually in the ``soft'' limit. Vertices of this type, however, would not mediate long--range HS effects, and actually can often be eliminated altogether by field redefinitions. Moreover, a recent analysis \cite{tar} that takes into account the explicit form of the cubic HS vertices of \cite{st,mmr} revealed that the problem presents itself \emph{only} when the exchanged particles have spins \emph{lower} than the external ones. We shall return to this important point in Section \ref{sec:beyondcubic}. However, the reader should already appreciate that, with these amplitudes alone, the possible long--range effects of HS particles would be confined to their own world, a scenario that was nicely foreseen by Porrati in \cite{porratiww}.

The next classic result, the Coleman--Mandula theorem \cite{cm}, is a formal $S$--matrix argument that has had a wide impact over the years. One of its key assumptions, however, the presence of finitely many HS excitations below any given mass level, makes it not directly applicable to the HS constructions one is after, whose gauge algebras bring about infinitely many massless gauge fields. The theorem does provide a rationale for the failure of naive attempts involving finitely many HS gauge fields, but I cannot help finding it less palatable and instructive than Weinberg's analysis.

The Velo--Zwanziger argument \cite{vz} can be regarded as a refinement of the old work of Fierz and Pauli \cite{dfp}. It concerns the loss of causality that HS fields tend to manifest in the presence of minimal couplings to uniform electromagnetic fields, and in this context String Theory has proved strikingly instructive. Indeed, Argyres and Nappi \cite{argnap} showed long ago that for the massive $s=2$ mode of the open string, that in some respects behaves like a HS field, the actual coupling to uniform electromagnetic fields, which rests on an exact conformal field theory and can thus be thoroughly analyzed, is free from such pathologies. The crux of the matter lies in the peculiar highly non--linear contact terms that are present in String Theory, and a recent extension of their analysis shows that all symmetric tensors of the first Regge trajectory work exactly in the same way, without any loss of causality \cite{prs}. However, the minimal couplings of these states to Electromagnetism fade out when the string tension or the HS masses tend to zero, so that HS interactions loose their infrared cutoff. The result resonates again with Weinberg's argument, and reinforces the feeling that HS fields do not like to be involved in \emph{long--range} interactions mediated by \emph{low--spin} particles.

The last classic results that I would like to describe point again in the same direction, albeit in different ways. They both indicate that HS fields \emph{do not} interact with gravity around flat space with two--derivative minimal couplings as their low--spin counterparts.

The Weinberg--Witten argument \cite{ww} compares two ways of determining the effect of rotations on matrix elements of the energy--momentum tensor $\langle f  | T_{\mu\nu}(q) | i \rangle$ between two spin--$s$ states in three dimensions. If $T_{\mu\nu}$ is assumed to behave as a Lorentz tensor, a property that it possesses when it is gauge invariant, and to involve only two derivatives, the argument shows that the matrix element vanishes for any type of \emph{massless} HS particles, be they fundamental or composite, and the authors present a similar argument for spin--one currents. The assumption is hardly a weak one, since already for $s=2$ the energy--momentum tensor $T_{\mu\nu}$ is \emph{not} gauge invariant, while already for $s=1$ spin--one currents are also not gauge invariant. Still, a recent refinement of \cite{ww} took into account the subtle behavior of $T_{\mu\nu}$ but reached nonetheless the same conclusion \cite{porratiww}. The problem has to do, once more, with the number of derivatives in HS couplings: if more derivatives were present, the matrix elements would vanish as $q \to 0$ and the arguments of \cite{ww,porratiww} would not apply.

Finally, the Aragone--Deser argument \cite{ad} rests on an explicit attempt to construct a standard gravitational coupling for a spin--$\frac{5}{2}$ field. The attempt fails, since the computation generates Weyl tensors, which cannot be compensated by the Einstein--Hilbert terms and thus constitute a genuine obstruction to the gauge symmetry. Similar results obtain for $s> 5/2$ Fermi fields or for $s>2$ Bose fields. This, however, is only true around flat space, and remarkably in the 1980's Fradkin and Vasiliev \cite{Fradkin} could show that the obstruction disappears in the presence of a cosmological term $\lambda$. In other words, in the presence of a cosmological term, and up to the cubic order, HS fields do allow conventional--looking gravitational couplings, albeit with an important proviso: they are inevitably accompanied by higher--derivative partners that, for dimensional reasons, bring along negative powers of $\lambda$. As a result, the solution to the Aragone--Deser problem found by Fradkin and Vasiliev has a singular behavior in the limit of a flat background.

The Aragone--Deser argument has had a profound influence on subsequent developments, which have been strongly biased toward the search for ``minimal--like'' HS interactions in the presence of a cosmological term. The main outcome of this research, the Vasiliev system \cite{vasiliev}, is remarkable and confusing at the same time. It is remarkable as a mathematical realization of free--differential algebras with non--polynomial scalar couplings and as an example of classically consistent HS interactions, and not surprisingly starts to play an important role in the AdS/CFT correspondence. It is confusing, however, since it points to intriguing dynamical effects. Vasiliev's fields are in fact a scalar, a vector and infinitely many symmetric tensors, one for each rank $s\geq2$, all valued in (anti)symmetric representations of a Chan--Paton group \cite{cp}. Hence, they are in one--to--one correspondence with the first Regge trajectory of the \emph{open} bosonic string. These fields have \emph{minimal}--like interactions, albeit around (A)dS backgrounds, and moreover the spin--2 singlet is naturally associated with gravity, while in String Theory gravity has to do with \emph{closed}, rather than open, strings. This circumstance would seem to suggest that a Cabibbo--like mixing takes place between closed and open fields, when they become degenerate in mass in the tensionless limit $\a^{\,\prime}\to \infty$. This point was raised in \cite{fms} and clearly deserves further attention. For all these reasons, and mainly to stress the key role of non--minimal interactions, both in the lectures that I have presented at Erice and in subsequent talks, I have kept a (slightly embellished) slide of my Strings 2009 presentation, where I had elaborated on the apparent tension between the Fradkin--Vasiliev minimal--like couplings and two recent results. One of these is due to Metsaev \cite{metsaev}, who completed the analysis that in the 1980's opened our first window on HS interactions \cite{goteborg}, and concerns the light--cone classification of cubic HS vertices in flat space, while the other is the subsequent scaling analysis of Boulanger, Leclercq and Sundell \cite{bls}. Briefly stated, Metsaev's analysis showed that cubic light--cone vertices exist around flat space in general, albeit with \emph{lower bounds} on the number of derivatives that depend on the collection of fields involved. For example, the $s=2$ couplings of a spin--$s$ field involve at least $2s-2$ derivatives. Boulanger, Leclercq and Sundell proposed a scaling argument that pulls out of the Fradkin--Vasiliev couplings the dominant contributions (what I called ``seeds'' in \cite{rome09}), and in so doing recovered explicit examples of covariant (as opposed to light--cone, as in Metsaev's analysis) cubic flat--space vertices. The existence of the scaling limit indicates that terms containing lower numbers of derivatives are really dressings due to the (A)dS background, no more no less than the mass--like terms that I shall revisit shortly. A later section will illustrate how String Theory sheds light on the issue. In the low--tension limit of its amplitudes, in fact, one can identify whole chains of covariant terms that fill Metsaev's list, whose high--derivative couplings naturally forego the previous restrictions.

Although this section is mostly devoted to the behavior around flat space, it actually begins with a brief discussion of how mass--like terms emerge in (A)dS backgrounds, starting from the free equations of Section \ref{sec:free}. I then turn to external currents and the van Dam--Veltman--Zakharov discontinuity  \cite{vdvz}, whose origin can now be related to special types of HS gauge symmetries that are allowed in dS backgrounds. External currents are also a first step toward more general Lagrangian couplings: the HS currents of free scalar fields, for instance, embody key information on their possible HS interactions and on their behavior at high energies \cite{bjm}. Section \ref{sec:cubicstringlessons} is then devoted to disk amplitudes for the open bosonic string and their neat lessons for the cubic interactions of symmetric Bose fields, and actually also of symmetric tensor--spinors \cite{st}. The end result of this analysis provides further clues that massive string spectra draw their origin from broken HS symmetries. The section ends with a brief review of some recent results of Taronna \cite{tar} that address explicitly the subtle behavior of the higher--point functions for massless HS fields in flat space.


\scss{Free HS fields in (A)dS backgrounds}\label{sec:adsdeformations}

As we have seen, the classic Aragone--Deser problem \cite{ad} is signalled by terms involving the Weyl tensor, which disappear in conformally flat backgrounds. This is an important exception that I cannot refrain from touching upon, since it underlies HS dynamics in the presence of a cosmological constant. The corresponding maximally symmetric backgrounds are (A)dS, rather than Minkowski, space times, where ordinary derivatives leave way to covariant derivatives such that
\be \left[ \, \nabla_\mu \, , \nabla_\nu \, \right] \, V_\rho \, =
\, \frac{1}{L^2} \left(  g_{\mu\rho} V_\nu \, - \, g_{\nu\rho} V_\mu
\right) \, . \label{covariantdev} \ee
For definiteness, here the sign is tailored to the dS case, $g$ is the background metric and $L$, the dS radius, is related to the cosmological constant in $D$ dimensions according to
\be
\lambda \, =\,  \frac{(D-1)(D-2)}{2\, L^2}\ . \label{adsradius}
\ee
Corresponding results for AdS obtain replacing $L^2$ with $-L^2$.

The commutator \eqref{covariantdev} brings about modifications of both ${\cal F}$ and its unconstrained completion ${\cal A}$ that make them compatible with the deformed gauge transformations
\be
\delta \, \varphi \, = \, \nabla \, \Lambda \ .
\ee
These modifications involve the replacement of ${\cal F}$ with
\be \cF_L \, = \, {\cal F} \, + \, \frac{1}{L^2} \, \left\{ \left[
(3-D-s)(2-s) - s \right]\, \varphi \ + \ 2 \, g \, \varphi^{'}
\right\} \, ,  \ee
where $D$ denotes the number of space-time dimensions and
\be {\cal F} \, = \, \Box \, \varphi \, - \, \nabla \, \nabla \cdot
\varphi \, + \, \nabla^{\, 2} \, \vf^{\, \pe}\,  \label{fl} \ee
is now built in terms of the covariant derivatives \eqref{covariantdev}. Notice that the deformed non--Lagrangian equations of motion, that are
now $\cF_L=0$, involve \emph{mass--like terms}. Mass--like terms, by no means ordinary mass terms, since they are precisely as needed for gauge invariance. At the same time
the gauge transformations of $\alpha$ and $\beta$ become
\be
\begin{split}
\d\, \a \,  &= \, \Lambda^{\, \pe} \ , \\
\d \, \b \, &= \, \nabla \cdot \nabla \cdot \nabla \cdot \Lambda \ .
\end{split}
\ee
Consequently, the $AdS$ deformations of $\cA$ and $\cC$ read
\be
\begin{split}
\cA_L \, & = \, \cF_L \, - \, 3 \, \nabla^{\, 3} \a \, - \,
\frac{4}{L^2}\ g \, \nabla \,  \a \, , \\
\cC_L \, & = \, \vf^{\, \pe \pe} \, - \, 4 \, \nabla \cdot \a \, - \, \nabla \, \a^{\, \pe} \, ,
\end{split}
\ee
and are rather simple, while
\be
\begin{split}
\cB_L \, & =  \, \b \, - \, \biggl\{\12 \, \nabla \cdot \nabla \cdot
\vf^{\, \pe} \, - \, \Box \, \nabla \cdot \a \, - \, \12 \, \nabla
\nabla
\cdot \nabla \cdot \a \\
            &  + \, \fr{1}{L^{\, 2}}\left( 2\, \nabla \a^{\, \pe} \,
+ \, 2\, g \, \nabla \cdot \a^{\, \pe} \, + \, \left[(s \, - \, 3)(5
\, - \, s \, - \, D)\right]\, \nabla \cdot \a \right) \biggr\} \
\end{split}
\ee
is more involved.

Let me stress that the actual form of the mass--like terms, which are mere curvature--induced dressings of the flat--space kinetic terms, is \emph{not} unique, but depends on the ordering chosen for the covariant derivatives. This is strikingly clear in the $s=1$ case, since the Maxwell equation admits the two naively different presentations
\be
\begin{split}
& \Box \, A_{\,\mu} \, - \, \nabla_{\, \nu}  \, \nabla^{\, \mu}\, A^{\, \nu}\, = \, 0 \ , \\
& \Box \, A_{\,\mu}\, - \, \nabla_{\, \mu}  \, \nabla \cdot A \, - \ \frac{D-1}{L^2} \ A_{\, \mu} \, = \, 0 \ ,
\end{split}
\ee
which are actually equivalent in  view of the commutator \eqref{covariantdev}.


\scss{External Currents and the vDVZ Discontinuity}\label{sec:extcurrents}

External currents are responsible for the simplest interactions in Field Theory, but much can be learned from them notwithstanding. Static sources give rise in this fashion to Coulomb's law or its HS generalizations, while generic sources encode important information on the number of propagating degrees of freedom.

One can start from the general Lagrangian field equations for symmetric tensors and extend them allowing for the presence of an external current ${\cal J}$, so that they become
\be
\begin{split}
& {\cal A}\, - \, \frac{1}{2} \, \eta \, {\cal A}^{\, \prime} \, { + \, \eta^2 \, {\cal B}} \, = \, {\cal J} \ ,   \\
& \partial \, \cdot {\cal A}^{\, \prime} { \,
- \,  (2\, \partial \, + \, \eta \, \partial \, \cdot)\, {\cal B}} \, = \, 0 \ , \\
& \varphi^{\prime \prime} { \, - \, 4 \, \partial \cdot \alpha \, - \, \partial \,
\alpha^{\, \prime}} \, = \, 0 \ ,\label{localspins}
\end{split}
\ee
where the unconstrained gauge symmetry demands that ${\cal J}$ be conserved. In the familiar $s=1$ case, in momentum space these equations reduce to
\be
 - \ p^2 \, A_\mu \ + \ p_\mu \, p \cdot A \ = \ {\cal J}_\mu \ ,
\ee
and the scalar product of this expression with the conserved current ${\cal J}_\mu$,
\be
{\cal E}_0(D,1) \, \equiv \, - \, p^2 \, {\cal J}^\mu \, A_\mu  \ = \  {\cal J}^\mu \, {\cal J}_\mu \ ,\label{exch1flat}
\ee
makes it possible, so to speak, to invert the kinetic operator without the need for gauge fixing, so that the ``current--exchange amplitude'' \eqref{exch1flat} identifies the residue of the pole. With an eye to the ensuing discussion, the subscript of ${\cal E}$ is meant to emphasize that the amplitude is computed in flat space, while its two arguments reflect the space--time dimension and the spin of the exchanged field. Notice that the square of the conserved current in \eqref{exch1flat} rests on $D-2$ independent contributions, as many as the independent polarizations of a massless spin--1 field in $D$ dimensions.

Before discussing the actual solution of the system \eqref{localspins}, let us elaborate briefly on the important lessons that this analysis has in store for the non--local formulation. In principle, the non--local formulation could be derived directly, integrating out both the compensator $\alpha$ and the Lagrange multiplier $\beta$ along the lines of \cite{francia10}, but current exchange amplitudes provide an alternative path that is conceptually important. The issue is selecting a non--local Lagrangian equation that yields the same physical effects, and thus the same current exchange amplitudes, as the local system \eqref{localspins}. Complete details can be found in \cite{fms}, but here we can give some clues that a proper non--local Lagrangian does indeed exist and is actually unique by taking a closer look at the $s=3$ case. In Section \ref{sec:freesymmetric} we identified two alternatives for the $s=3$ field, and now we can see explicitly how external currents can distinguish the pseudo--differential operators ${{\cal F}^{(1)}}$ and ${\widetilde{\cal F}^{(1)}}$ of eqs.~\eqref{s3nl1} and \eqref{s3nl2}. Their differences reflect themselves in their Bianchi identities, that in index--free notation read
\be
\begin{split}
& \partial \cdot {{\cal F}^{(1)}} \, - \, \frac{1}{4} \ \partial \, {{\cal F}^{(1)}}^\prime \, = \, 0 \ ,\\
& \partial \cdot {\widetilde{\cal F}}^{(1)} \, - \, \frac{1}{2} \ \partial \, {\widetilde{\cal F}}^{(1)
\,^\prime} \, = \, 0 \ ,
\end{split}
\ee
where the first is a special case of eq.~\eqref{bianchin} while the second can be computed directly  and
coincides with the Bianchi identity for the Fronsdal operator ${\cal F}$. As a result, in the two cases one is led to define the divergence--free Einstein--like tensors
\be
{\cal G}^{(1)} \, = \, {{\cal F}^{(1)}}\, -\, \frac{1}{4} \ \eta \, {{\cal F}^{(1)}}^\prime \qquad {\rm and} \qquad
\widetilde{\cal G}^{(1)} \, = \, {\widetilde{\cal F}^{(1)}}\, -\, \frac{1}{2} \ \eta \, {\widetilde{\cal F}^{(1)\, ^\prime}} \ ,
\ee
and the corresponding Lagrangian equations, where they are sourced by conserved external currents, then yield the current exchange amplitudes
\be
{\cal J} \cdot {\cal F}^{(1)} \, \equiv {\cal J} \cdot \Box \varphi \,= \, {\cal J} \cdot {\cal J} \, - \, \frac{3}{D-2} \, \ {\cal J}^\prime \cdot {\cal J}^\prime
\ee
and
\be
{\cal E}_0(D,3)\, \equiv \, {\cal J} \cdot \widetilde{\cal F}^{(1)} \, \equiv {\cal J} \cdot \Box \varphi \, = \, {\cal J} \cdot {\cal J} \, - \, \frac{3}{D} \, \ {\cal J}^\prime \cdot {\cal J}^\prime \ .
\ee
$\widetilde{\cal G}^{(1)}$ is the correct Einstein--like tensor in this case, since one can verify that only the last expression identifies a traceless and transverse $s=3$ current.

The detailed analysis in \cite{fms} provides a clear explanation of how the Bianchi identities for the ${\cal F}^{(n)}$ operators of Section \ref{sec:freesymmetric} determine the proper non--local Einstein--like tensors. For any given $s$, for $n$ large enough these operators become gauge invariant, while their Bianchi identities loose memory of the term proportional to $\vf^{\,\prime\prime}$ and simply relate divergences to gradients of their traces. Therefore, combining traces of the first gauge--invariant ${\cal F}^{(n)}$ with corresponding powers of $\frac{\partial^{\,2}}{\Box}$ one ought to arrive at a non--local analogue of the gauge--invariant $\cA$ tensor of Section \ref{sec:freesymmetric}. As shown in \cite{fms}, this is indeed true, and moreover there is a unique choice to this effect. Consequently, the non--local Lagrangians make full use of the independent gauge--invariant operators that are built by eq.~\eqref{iteration}, as we had stated in Section \ref{sec:freesymmetric}, and the end results yield the same current exchange amplitudes as the local theories. Here we can content ourselves with a brief discussion of the final result,
\begin{eqnarray}
&& {\cal E}_0(D,s) \, \equiv \, {\sum_{n=0}^{N}\ \rho_n(D-2,s) \ \frac{s!}{{n!\; (s-2n)!\; 2^n}}\
{\cal J}^{[n]}\cdot {\cal J}^{[n]}} \ , \label{exchangeD} \\
&& { \rho_{n+1}(D,s)\ =\ -\  \frac{\rho_n(D,s)}{ D+2(s-n-2)}} \ , \qquad \rho_0(D,s)\, =\, 1 \ , \label{recursionD}
\end{eqnarray}
which can be justified noticing that the recursive definition \eqref{recursionD} of the $\rho_n$ makes the sum \eqref{exchangeD} transverse and traceless, as in the preceding $s=3$ example. The first non--trivial case corresponds to $s=2$, where eq.~\eqref{exchangeD} reduces to
\be
{\cal E}_0(D,2) \, = \, T^{\mu\nu}\, T_{\mu\nu} \ - \ \frac{1}{D-2} \ \left( {T^{\,\mu}}_\mu\right)^2 \label{exchangeDs2}
\ee
where, abiding to standard conventions, we have called $T_{\mu\nu}$ the corresponding current.

Let me stress that the amplitude ${\cal E}_0(D,s)$ depends on the spin $s$ of the HS field determining the exchange and on the number $D$ of space--time dimensions. If one insists on currents that are \emph{conserved} in $D$ dimensions, it is possible
to adapt the analysis to \emph{massive} spin--$s$ fields. A harmonic dependence on an
internal circle coordinate suffices in fact to introduce masses \emph{\`a la} Stueckelberg, so that massive exchanges driven by currents that are still \emph{conserved} in $D$ dimensions and lack internal components obtain replacing $D$ with $D+1$ in eq.~\eqref{exchangeD}. The difference between the exchanges for any given value of $s$ computed in $D+1$  and $D$ dimensions,
\be
\Delta_0(D,s) \, = \, {\sum_{n=0}^{N}\ \left[ \rho_n(D-1,s)\, - \, \rho_n(D-2,s)\right] \ \frac{s!}{{n!\; (s-2n)!\; 2^n}}\
{\cal J}^{[n]}\cdot {\cal J}^{[n]}}\ ,
\ee
thus encodes the generalization of the discontinuity originally found by van Dam, Veltman and Zakharov \cite{vdvz} for $s=2$, that can be computed starting from eq.~\eqref{exchangeDs2}.

This phenomenon has a very instructive counterpart in the presence of a cosmological constant $\lambda$ that deforms Minkowski space to (A)dS backgrounds depending on its (negative)positive sign. Starting from the deformed equations of Section \ref{sec:adsdeformations} and focussing on the dS case, for $s=2$ the resulting massive exchange for a generic mass $M$, computed some time ago in \cite{hp}, reads\footnote{This mass deformation should not be confused with the special mass parameters of Section \ref{sec:adsdeformations}, which guarantee the massless gauge symmetry in (A)dS backgrounds.}
\be
{\cal E}_{\lambda}(D,2)\, = \, T^{\mu\nu}\, T_{\mu\nu}  \, - \,  \frac{1}{D-1}\
\frac{(ML)^2-(D-1)}{(ML)^2-(D-2)}\ \left({T^\mu}_\mu \right)^2 \ , \label{elambda2}
\ee
where the relation between $L$ and the cosmological constant is given in eq.~\eqref{adsradius}. Notice that this exchange is a rational function of $ML$ with a simple pole determined by the condition
\be
(ML)^2 = D-2 \ . \label{pmass}
\ee

The pole in eq.~\eqref{elambda2} is a manifestation of an interesting phenomenon, a shortening that occurs in dS representations precisely for the value \eqref{pmass} of $(ML)^2$ \cite{shortening}. Indeed, as first noticed in \cite{pgauges}, when eq.~\eqref{pmass} holds the theory acquires the ``partial gauge symmetry''
\be
\delta \varphi_{\mu\nu} = \nabla_{\mu}\nabla_{\nu} \, \zeta \,+\, \frac{M^2}{D-2}\ g_{\mu\nu} \, \zeta \ , \label{pgauget}
\ee
with $\zeta$ a scalar parameter.

There is a marked difference between eq.~\eqref{pgauget} and the standard massless gauge symmetry
\be
\delta \varphi_{\mu\nu} = \nabla_{\mu}\, \xi_{\, \nu} \,+\, \nabla_{\nu}\, \xi_{\, \mu} \ ,
\ee
with $\xi_\mu$ a vector parameter, since the coupling to a generic conserved energy--momentum tensor is \emph{not} compatible with the second term in eq.~\eqref{pgauget} unless the trace of $T_{\mu\nu}$ vanishes identically. This fact also brings about the vDVZ discontinuity, since the limiting values of the rational function in eq.~\eqref{elambda2} for $ML\to 0$, that identifies the flat massless case, and for $ML \to \infty$, that identifies the flat massive case, differ in compliance with Liouville's theorem. In other words, the vDVZ discontinuity in flat space is somehow a reflection of the partial gauge symmetry that emerges, for real fine--tuned values of $(ML)^2$, in dS space.

More details can be found in \cite{fms}, where eq.~\eqref{elambda2} was extended to symmetric tensors of arbitrary rank. The first terms, drawn from \cite{fms} but adapted to the index--free notation of Section \ref{sec:freesymmetric}, read
\be
\begin{split}
&{{\cal E}_\lambda(D,s)}\ {=}\ {\cal J}_s\ +\ {{g \,{\cal J}_s^{\;\prime}} \over
2(\frac{5}{2}- \zeta)}\ {(ML)^2+2(\frac{5}{2}-\zeta)\over
(ML)^2-2(\zeta-3)}\ \\ {+}\ &{{g^2 \, {\cal J}_s^{\;[2]}} \over
8}\ {(ML)^4+8(ML)^2(\frac{7}{2}-\zeta)+12\left({5\over
2}-\zeta\right)_2
\over\left({5\over
2}-\zeta\right)_2[(ML)^2-2(\zeta-3)][(ML)^2-6(\zeta-4)]} \label{spinsds}\\
{+} \ &{{g^3 \, {\cal J}_s^{\;[3]}} \over 48}\ {(ML)^6-(ML)^4
(18\zeta-77)+92(ML)^2\left({7\over
2}-\zeta\right)_2+120\left({5\over
2}-\zeta\right)_3\over\left({5\over
2}-\zeta\right)_3[(ML)^2-2(\zeta-3)][(ML)^2-6(\zeta-4)]
[(ML)^2-10(\zeta-5)]} \\
{+}\ &{\dots } \ ,
\end{split}
\ee
where ${\cal J}_s^{\;[2,3]}$ are higher traces of ${\cal J}_s$, $g$ is the background dS metric, $\zeta = \frac{D}{2}+s$ and
\be
(a)_n=a(a+1)\ldots (a+n-1)
\ee
are Pochhammer symbols.

In the dS case the poles lie precisely at real values of $ML$ where partially massless gauge transformations involving terms like the second in eq.~\eqref{pgauget}, not protected by gradients and thus incompatible with generic conserved currents, emerge. For a rank--$s$ tensor there are in fact $s-1$ partially massless points, for which
\be
 (ML)^2=(s-1-r)(D+s+r-4)\, ,
\ee
where $r=0,\ldots,s-2$, while $r=s-1$ corresponds to the more familiar massless point. For all $s\geq 2$ a first pole thus appears at $r=s-2$, and for $s>3$ others lie at alternate values of $r$ below it. For instance, for $s=3$ the relevant value is $r=1$, since for $r=0$ all terms present in the partially massless gauge transformation contain gradients of the parameter and are thus manifestly compatible with conserved currents.

One can derive eq.~\eqref{spinsds} starting from a $D+1$--dimensional flat Minkowski space and performing a radial reduction \cite{radial}, a generalization of the polar decomposition of Euclidean space. Among the technical complications discussed in \cite{fms}, I would like to mention here that the end result was recast in a form that depends manifestly on $(ML)^2$ via interesting identities for the generalized hypergeometric functions $_3F_2$, more complicated cousins of the familiar $_2F_1$ \cite{whiwat}.

Let me stress that a close analogy exists between the singular behavior of dS exchanges at partially massless points and the massless limit of the more familiar flat--space exchanges in the presence of non--conserved currents. For instance, in the Proca theory a non--conserved current would lead to the exchange
\be
\left( p^{\, 2} \, + \, m^{\, 2} \right) J \cdot A \, = \, J \cdot J - \frac{1}{m^2} \, \left(p \cdot J\right)^2 \ ,
\ee
whose massless limit is singular precisely because a current that is not conserved conflicts with the emerging gauge symmetry. Following \cite{vdvz}, we insisted on conserved currents also in the massive case, and this allowed a direct comparison between massive and massless exchange amplitudes.

A further step along these lines was made by Bekaert, Joung and Mourad in \cite{bjm}. The key idea of their work was to apply the current--exchange formula \eqref{exchangeD} to the \emph{dynamical} Noether HS currents of a complex scalar field $\Phi$. These can be defined, as in \cite{bbvd}, via the Wigner function
\be
 J(x,v)= \bar{\vf}(x+i v) \, \vf(x - i v) \, , \label{scalarHScurrent}
\ee
where the spin--$s$ term is combined with $s$ powers of the vector $v$. At the same time, the strength of the interactions with generic HS symmetric tensors can be characterized via a single dimensionful coupling, that by analogy with String Theory I will call $\alpha^{\, \prime}$ in the following, and via an infinite number of dimensionless couplings $a_k$, or rather via a \emph{coupling function}
\be
a(z)=\sum_r \frac{z^r}{r!}\ a_r \ , \label{couplingfunction}
\ee
that in String Theory would be dominated by an exponential.
Classic results in the theory of orthogonal polynomials allow one to combine the current--exchange amplitude \eqref{exchangeD}, a pair of HS currents as in eq.~\eqref{scalarHScurrent} and the coupling function \eqref{couplingfunction} into a compact formula for
the interaction between complex scalars and infinitely many massless HS fields. In four dimensions the result is strikingly neat, and for the $\vf+\vf \to \vf+\vf$ amplitude reads
\be
\begin{split}
{\cal A}^{(s)} =\, -\, &\frac{1}{\alpha^{\,\prime}s}\left[a\left(\frac{\alpha^{\,\prime}}{4}\,(u-t)+\frac{\alpha^{\,\prime}\!\!}{2}\, \sqrt{-ut}\right)+ a\left(\frac{\alpha^{\,\prime}}{4}\,(u-t)-\frac{\alpha^{\,\prime}\!\!}{2}\,\sqrt{-ut}\right)-a_0\right]\ \times \\
& \varphi_{\,1}\left(p_{\,1}\right) \varphi_{\,2}\left(p_{\,2}\right) \varphi_{\,3}\left(p_{\,3}\right)
\varphi_{\,4}\left(p_{\,4}\right) \ , \label{bjmformula}
\end{split}
\ee
where $s$, $t$ and $u$ are the familiar Mandelstam variables, since the sum contributing to the current exchange can be related to the Chebyshev polynomials \cite{whiwat}.

This beautiful expression has a number of interesting lessons in store. For one matter, it is a consistent four--scalar amplitude involving the exchange of infinitely many \emph{massless} HS particles. Moreover, the detailed discussion in \cite{bjm} shows that, in principle, a soft behavior at high energies can be attained working only with (infinitely many) symmetric fields, provided the coupling function tends to zero for large negative real values of its argument. In String Theory the essential singularity of $a(z)$ may be held ultimately responsible for the presence of lower Regge trajectories, since a soft behavior for the conjugate amplitude $\vf+\bar{\vf} \to \vf+\bar{\vf}$ would also demand that $a(z)$ tend to zero for large positive real arguments.  Therefore, as stressed in \cite{bjm}, in the present setting a soft behavior for all conjugate amplitudes would require that the coupling function $a(z)$ tend to zero at infinity in the complex plane. This is a subtle condition, since Liouville's theorem would then require the presence of singularities in the finite plane, which in their turn would signal in general an extended nature for the objects involved. There is clearly more to be understood here, and other intriguing properties will show up in the ensuing discussion.


\scss{String lessons for cubic HS couplings}\label{sec:cubicstringlessons}

I can now turn to the scattering amplitudes of the open bosonic string involving its massive HS excitations. As we shall see, they contain a wealth of information on HS couplings \cite{st}, and this analysis provides some further evidence for the long--held, but never fully quantified, expectation that String Theory describes a broken phase for HS gauge symmetries. This brings about a natural link between cubic couplings and Noether currents, with the potential of clarifying also the nature of the latter. The situation is particularly manageable for the symmetric tensors of the first Regge trajectory, whose cubic disk amplitudes can be simply computed from the path integral for the free action
\be
S_P[X,\g]\,=\,-\, \frac{1}{4\pi\a^{\,\prime}}\int_M d^{\,2} \xi \, \eta^{\,ab}\,\partial_{\,a} X^\m \, \partial_{\,b} X_\m  \label{Polyakov}
\ee
in the presence of special external currents. Eq.~\eqref{Polyakov} is in fact the remnant of the bosonic string action after a complete gauge fixing in the critical dimension, while the vertex operators of the first Regge trajectory, combinations of an exponential with powers of $\partial X^{\, \mu}$, can all be recovered as Taylor coefficients in the ``symbols" $\xi^{\, \mu}$ of
\be
\cV(\sigma,p,\xi) \, = \, \exp\bigg[i p \cdot X(\sigma) \, +\,  \xi \cdot \partial X(\sigma) \bigg] \ . \label{vertex_general}
\ee

Up to a measure factor that can be ascribed to ghost fields, tree--level correlation functions of these types of vertices on the disk are then tantamount to a gaussian path integral in the presence of the boundary currents
\begin{equation}
J(\sigma, p, \xi)\,=\,\sum_{i\,=\,1}^N\left(p_{\,i}\,\delta^{\,2}(\s-\s_i)\,
- \, \frac{{\xi}_{\,i}}{\sqrt{2\a^{\,\prime}}}
\ \partial_\sigma \, \delta^{\,2}(\s-\s_i)\right)\ .\label{curr}
\end{equation}
All these amplitudes, and three--point functions in particular, can be extracted from
\be
\begin{split}
S^{open}_{j_1\cdots j_n}=\int_{\mathbb{R}^{n-3}}\ dy_4\cdots &dy_{n}\ |y_{12}y_{13}y_{23}| \ \times\\ & \bra\cV_{j_1}(\hat{y}_1)\cV_{j_2}(\hat{y}_2)\cV_{j_3}(\hat{y}_3)\cdots\cV_{j_n}({y}_n)\ket Tr(\L^{a_1}\cdots\L^{a_n}) \ , \label{diskamplitude}
\end{split}
\ee
where the trace is a Chan--Paton factor \cite{cp} and the integrals are computed along the real axis, so that the interior of the disk corresponds to the upper half--plane. For brevity, we have left implicit all momenta and $\xi$--variables for the $n$ vertex operators in \eqref{diskamplitude}. Standard field theory techniques then imply that correlation functions of the vertex operators, determined by the two--dimensional Green function, take the form
\be
\begin{split}
\bra\cV_{j_1}(\hat{y}_1) \cdots \, \cV_{j_n}({y}_n)\ket & =  \\ & \exp\left[\sum_{i\neq j}^N\left(\a^{\,\prime}p_{\,i}\cdot p_{\,j}
\,\ln|y_{ij}|+\sqrt{2\a^{\,\prime}}\ \frac{\xi_{\,i}\cdot p_{\,j}}{y_{ij}}+\frac{1}{2}\ \frac{\xi_{\,i}
\cdot \xi_j}{y_{ij}^2}\right)\right]\ ,\label{Gen3}
\end{split}
\ee
where the $y_i$ denote the locations of the punctures along the real axis and $y_{ij}=y_i-y_j$.

Eq.~\eqref{Gen3} would seem to violate basic tenets of String Theory, since the dependence on the location of the punctures ought to disappear from three--point functions, up to a measure factor compensating exactly the one introduced by ghosts in eq.~\eqref{diskamplitude}. Of course, this property ought to hold solely for on--shell external states satisfying the Virasoro conditions, so that a closer look into the matter is necessary before coming to definite conclusions. The independent Virasoro conditions,
\be
\begin{split}(L_0-1)\,\left|\Psi\right\rangle &= 0  \ , \\
L_1\,\left|\Psi\right\rangle &= 0  \ , \\
L_2\,\left|\Psi\right\rangle &= 0
\end{split}
\ee
translate directly into the DFP conditions \eqref{DFP1} of Section 2, as the reader can simply verify, and considering three external states such that
\begin{align}
-p_{\,1}^{\,2}\,=\,&\frac{n_1-1}{\a^{\,\prime}}\ ,& -p_{\,2}^{\,2}\,=\,&\frac{n_2-1}{\a^{\,\prime}}\ ,& -p_{\,3}^{\,2}\,=\,&\frac{n_3-1}{\a^{\,\prime}}\ ,
\end{align}
the three--point correlation functions \eqref{Gen3} take the form
\be
\left|\frac{y_{12}y_{13}}{y_{23}}\right|^{n_1}
\left|\frac{y_{12}y_{23}}{y_{13}} \right|^{n_2}\left|\frac{y_{13}y_{23}}{y_{12}}\right|^{n_3} \\\times\exp\left[\sum_{i\neq j}^3\left(\frac{1}{2}\frac{\xi_{\,i}\cdot \xi_{\,j}}{y_{ij}^{\,2}}
+\sqrt{2\a^{\,\prime}}\ \frac{\xi_{\,i}\cdot p_{\,j}}{y_{ij}}\right)\right] \ . \label{cubicstring}
\ee

Apparently, this expression still depends on the location of the punctures. However, on account of the Virasoro conditions, the spins of the external states should be correlated to their mass levels, so that in eq.~\eqref{cubicstring} one should only retain terms containing $n_1$ powers of $\xi_1$, $n_2$ powers of $\xi_2$ and $n_3$ powers of $\xi_3$. It is then pleasing to verify how, for this class of terms, the dependence on the $y_i$ disappears altogether, so that one is finally left with
\be
\begin{split}
\exp\biggl\{\sqrt{\frac{\a^{\,\prime}\!\!}{2}}\, \bigl[\xi_{\,1}\cdot
p_{\,23}\,&\left\langle\frac{y_{23}}{y_{12}y_{13}}\right\rangle + \,\xi_{\,2}\cdot
p_{\,31}\,\left\langle\frac{y_{13}}{y_{12}y_{23}}\right\rangle\\  &+\,\xi_{\,3}\cdot
p_{\,12}\,\left\langle\frac{y_{12}}{y_{13}y_{23}}\right\rangle\bigr]\ +\ \bigl[\xi_{\,1}\cdot
\xi_{\,2}+\xi_{\,1}\cdot \xi_{\,3}+\xi_{\,2}\cdot \xi_{\,3}\bigr]\vphantom{\sqrt{\frac{\a^{\,\prime}\!\!}{2}}}\biggr\} \label{Zphys} \ .
\end{split}
\ee
In this expression the brackets are merely signs: they reflect the flip symmetry \cite{cp} of string amplitudes, which is implemented by projective--disk amplitudes and plays a crucial role whenever external states of odd spin are present.

Eq.~\eqref{Zphys} should be translated into the language of Field Theory, taking into account that $s$ powers of the $\xi_i$ are meant to correspond to a spin--$s$ field of the first Regge trajectory, a rank--$s$ symmetric tensor. Alternatively, one would like to associate to the external fields generic polarization tensors rather than simply powers of the $\xi_i$'s, and this is effected by a $\star$ -- product, a convenient procedure to multiply pairs of series. Thus, if
\be
A(\xi)\,=\, \sum_{k=0}^\infty A_{\mu_1 \ldots \, \mu_k}\, \frac{\xi^{\, \mu_1}\ldots\,  \xi^{\, \mu_k}}{k!}
\label{Axi}
\ee
and
\be
B(\xi)\,=\, \sum_{k=0}^\infty B_{\mu_1 \ldots \, \mu_k}\, \frac{\xi^{\, \mu_1}\ldots\,  \xi^{\, \mu_k}}{k!}\ ,
\ee
then
\be
A \star B \,=\, \sum_{k=0}^\infty \frac{1}{k!} \ A^{\mu_1 \ldots \, \mu_k} \, B_{\mu_1 \ldots \, \mu_k}\, ,
\ee
which is indeed reconstructing generic polarization tensors from the special low--rank tensors determined by the generating function \eqref{Gen3}. The standard presentation of this $\star$ -- product is
\be
A \star B \, = \, \left. \exp\left( \frac{\partial}{\partial \xi} \cdot \frac{\partial}{\partial \eta} \right) A(\xi) \,B(\eta) \right|_{\xi=\eta=0} \ ,
\ee
while a second presentation,
\be
A \star B \, = \, \int \frac{d^d\, U}{(2\pi)^{\,d/2}}\ \tilde{A}(U)\,B(i\, U)\ ,
\ee
where $\tilde{A}$ denotes the Fourier transform of $S$ with respect to the symbols, follows inserting into the first the definition of $A$ via an inverse Fourier transform. More details on how these results lead directly to the general cubic couplings
\be
\left.{\varphi}_{\,1}\left(p_{\,1},{\partial_\xi}\pm\sqrt{\frac{\a'}{2}}\,p_{\,31}\right)\,
{\varphi}_{\,2}\left(p_{\,2},\xi+{\partial_\xi}\pm\sqrt{\frac{\a'}{2}}\,p_{\,23}\right)\,
{\varphi}_{\,3}\left(p_{\,3},\xi\pm\,\sqrt{\frac{\a'}{2}}\,p_{\,12}\right) \
\right|_{\xi=0}  \label{cubicHS} \ee
can be found in \cite{st}. This expression is to be computed at $\xi=0$, $p_{ij}=p_i-p_j$ and the notation is as in eq.~\eqref{Axi}, so that for example the first factor is
\be \sum_{k=0}^\infty \varphi_{\mu_1 \ldots \, \mu_k}\left(p_{\,1}\right) \, \frac{\left({\partial_\xi}\pm\sqrt{\frac{\a'}{2}}\,p_{\,31}\right)^{\, \mu_1}\ldots\,  \left({\partial_\xi}\pm\sqrt{\frac{\a'}{2}}\,p_{\,31}\right)^{\, \mu_k}}{k!} \ ,
\ee
where derivatives act to the right. For traceless $\vf$ fields there is no ordering ambiguity.

It is instructive to expand eq.~\eqref{cubicHS} for the first few cases,
\be
\begin{split}
{\cal A}^{\,\pm}_{\,0-0-s}&=\left(\!\pm\sqrt{\frac{\alpha^{\,\prime}\!\!}{2}}\,\right)^{s}\varphi_{\,1}\,\varphi_{\,2}\ \varphi_{\,3}\cdot
p_{\,12}^{\,s}\ , \\
{\cal A}^{\,\pm}_{\,1-1-s}&=\left(\!\pm\sqrt{\frac{\alpha^{\,\prime}\!\!}{2}}\,\right)^{s-2}\,s(s-1)\,
A_{1\,\mu}\,A_{2\,\nu}\,\varphi^{\,\mu\nu\ldots}\,p_{\,12}^{\,s-2} \\
&+\left(\!\pm\sqrt{\frac{\alpha^{\,\prime}\!\!}{2}}\,\right)^s \Big[A_{1} \cdot A_{2}\ \varphi\cdot
p_{\,12}^{\,s}+s\,A_{1}\cdot p_{\,23}\,A_{2\,\nu}\,\varphi^{\,\nu\ldots\,}p_{\,12}^{\,s-1} \\
&+ s\,A_{2}\cdot p_{\,31}\, A_{1\,\nu}\,\varphi^{\,\nu\ldots\,}p_{\,12}^{\,s-1}\Big] \\
&+\left(\!\pm\sqrt{\frac{\alpha^{\,\prime}\!\!}{2}}\,\right)^{s+2}A_{1} \cdot
p_{\,23}\,A_{2}\cdot p_{\,31}\, \varphi\cdot p_{\,12}^{\,s}\ , \label{11s}
\end{split}
\ee
the first of which corresponds to the Wigner currents \eqref{scalarHScurrent} used in \cite{bjm}.
Notice that the $0-0-s$ amplitude in \eqref{11s} does not retain any memory, so to speak, of the fact that for $s>1$ the $\vf$ fields are \emph{massive} in the open bosonic string. Indeed, if all momenta were those of massless fields it would be automatically invariant under the gauge transformation \eqref{gaugessred} up to Virasoro or DFP conditions, which would imply $p_i \cdot p_j = 0$ for all $i$ and $j$. On the other hand, even with all massless momenta the first line of the $1-1-s$ amplitude, which contains its lowest--derivative terms, would be incompatible with the spin--$s$ gauge symmetry.

This feature of cubic vertices is along the lines of what we have stressed in preceding sections: in the free string spectrum, mass generation \emph{\`a la} Stueckelberg rests precisely on terms that interactions feed into the quadratic level, consistently with the fact that a conventional gauge symmetry is recovered if these are left aside. This is what happens in the limit $\a^\prime \to\infty$, \emph{i.e.} in the limit of vanishing string tension. Hence, if a breaking phenomenon were really at work in String Theory, contributions from higher orders ought to pop up amidst cubic couplings. This phenomenon is strikingly visible at the cubic level: once the non--invariant terms, that carry along higher powers of the order parameter $\frac{1}{\sqrt{\a^\prime}}$ for the breaking, are taken apart, they leave behind a conventional gauge symmetry. Unfortunately, $\a^\prime$ lacks a dynamical origin in our present formulation of String Theory, while a minor subtlety is that all this occurs up to divergences and traces in these S--matrix amplitudes. Still, any gauge variations of the cubic vertices that are proportional to $p_{i}^{\,2}$ signal the mere need for deformed gauge transformations, up to trivial redefinitions, since any non--linear gauge variations of the kinetic terms that could in principle compensate them would be inevitably proportional to the free field equations.

Let us therefore remove from the cubic vertices all terms like the first in eq.~\eqref{11s} and focus, for all values of $s_1$, $s_2$ and $s_3$, on contributions that become gauge invariant on shell when the masses of the external states are removed, The end result, rather rewarding and strikingly simple, is captured by the expression  \cite{st} \footnote{Let me stress that retaining all these gauge--invariant terms is a weaker condition than simply letting $\a^\prime\to \infty$, which would only leave the leading terms with $s_1+s_2+s_3$ derivatives that I shall describe shortly. The leading terms, incidentally, were first identified long ago by Gross and Mende \cite{grossmende}.}
\be
\cA^{\,\pm}\,=\, e^{\pm\, \Gamma} \  \varphi_{\,1}\left(p_{\,1},\,\xi_{\,1}\right)\,
\varphi_{\,2}\left(p_{\,2},\,\xi_{\,2}\right)\,
\varphi_{\,3}\left(p_{\,3},\,\xi_{\,3}\right)\Bigg|_{\xi_{\,i}\,=\,0} \ , \label{cubicmassless}
\ee
so that it rests on the operator
\be
\begin{split}
\Gamma \, = \sqrt{\frac{\a'}{2}}\Big\{&\left[1+(\partial_{\xi_{\,1}}\cdot\partial_{\xi_{\,2}})\right](\partial_{\xi_{\,3}}\cdot p_{\,12})\,+\,\left[1+(\partial_{\xi_{\,2}}\cdot\partial_{\xi_{\,3}})\right]
(\partial_{\xi_{\,1}}\cdot p_{\,23}) \\ & +\,\left[1+(\partial_{\xi_{\,3}}\cdot\partial_{\xi_{\,1}})\right](\partial_{\xi_{\,2}}\cdot p_{\,31}) \Big\} \ ,
\label{Gamma}
\end{split}
\ee
which commutes with gauge transformations (in this notation, with terms of the form $p_i \cdot \xi_i$), up to Virasoro or DFP conditions.  There is therefore a unique term containing $s_1+s_2+s_3$ powers of momenta, proportional to
\be
p_{23}^{s_1} \cdot \varphi_1(p_1) \ p_{31}^{s_2} \cdot \varphi_2(p_2) \ p_{12}^{s_3} \cdot \varphi_3(p_3) \ ,
\ee
together with other groups of terms containing lower numbers of derivatives, whose structure is best captured by an equivalent expression for $\cA$,
\be
\begin{split}
\cA^{\,\pm}\,=\,\vphantom{\int}e^{\, \pm \, \cG} \ & \phi_{\,1}\left(p_{\,1}\,,\xi_{\,1} \pm \sqrt{\frac{\a^{\,\prime}\!\!}{2}}\,p_{\,23}\right)\ \times \label{cubicmassless2} \\ & \left.
\phi_{\,2}\left(p_{\,2}\, ,\xi_{\,2} \pm \sqrt{\frac{\a^{\,\prime}\!\!}{2}}\,p_{\,31}\right)\, \phi_{\,3}\left(p_{\,3} \, ,\xi_{\,3}\pm\sqrt{\frac{\a^{\,\prime}\!\!}{2}}\,p_{\,12}\right)\ \right|_{\xi_{\,i}\,=\,0} ,
\end{split}
\ee
where $p_{ij}=p_i-p_j$, the expression is to be evaluated at ${\xi_{\,i}\,=\,0}$ and
\be
\cG \, =\, \sqrt{\frac{\a'}{2}}\Big[(\partial_{\xi_{\,1}}\cdot\partial_{\xi_{\,2}})(\partial_{\xi_{\,3}}\cdot p_{\,12})\,+\,(\partial_{\xi_{\,2}}\cdot\partial_{\xi_{\,3}})
(\partial_{\xi_{\,1}}\cdot p_{\,23})\,+\,(\partial_{\xi_{\,3}}\cdot\partial_{\xi_{\,1}})(\partial_{\xi_{\,2}}\cdot p_{\,31}) \Big] \ .
\label{GammaG}
\ee

Notice that $\cG$ reduces indeed by two units the total number of momenta and removes at the same time three $\xi_i$'s, one from each of the fields. Therefore if $s_3$, say, is the smallest of the three spins, the operation can be repeated at most $s_3$ times, leading precisely to the removal of $2 \,s_3$ of the original momenta. In particular, for three spin--1 external states eqs.~\eqref{cubicmassless} and \eqref{cubicmassless2} yield both the three--derivative vertex displayed in Schwarz's 1982 review \cite{cp} and the cubic Yang--Mills vertex. Similarly, for three spin--2 external states they yield three groups of terms with six, four and two derivatives, the last of which is the standard cubic Einstein vertex. As a result, eqs.~\eqref{cubicmassless} and \eqref{cubicmassless2} involve terms with overall powers of momenta ranging between $s_1+s_2+s_3-2 \,{\rm min} \{s_i\}$ and $s_1+s_2+s_3$, precisely as in Metsaev's light--cone classification of cubic HS interactions \cite{metsaev}.\footnote{I cannot fail to mention that a very recent (A)dS extension \cite{adsembedding} shows in a concise and elegant fashion how the covariant derivatives \eqref{covariantdev} generate Fradkin--Vasiliev minimal--like dressings, the counterparts for cubic vertices of the mass--like terms of Section \ref{sec:adsdeformations}, although this goes well beyond the Erice School.}

Infinitely many gauge--invariant three point functions are tantamount to infinitely many conserved currents
that generalize the Wigner function of eq.~\eqref{scalarHScurrent},
\be
\begin{split}
{\cal J}^\pm(&x\,;\,\xi)\,=\,
\exp\left(\mp i\sqrt{\frac{\a^{\,\prime}\!\!}{2}}\ \xi_{\,\a}
\left[\partial_{\zeta_1}\cdot\partial_{\zeta_2}\,\partial^{\,\a}_{\,12}-2\,\partial^{\,\a}_{\zeta_1}\,
\partial_{\zeta_2}\cdot \partial_{\,1}
+\,2\,\partial^{\,\a}_{\zeta_2}\,\partial_{\zeta_1}\cdot \partial_{\,2}\right]\right) \ \times \label{bosecurents}\\
&\,\varphi_{\,1}\left(x_1 \mp i\sqrt{\frac{\a^{\,\prime}\!\!}{2}}\ \xi ,\zeta_1 \mp i \sqrt{2\a^{\,\prime}} \partial_{\,2}\right)\,
\varphi_{\,2}\left(x_2 \pm i\sqrt{\frac{\a^{\,\prime}\!\!}{2}} \xi , \zeta_2 \pm i \sqrt{2\a^{\,\prime}} \partial_{\,1}\right)\Bigg|_{\zeta_i\,=\,0} ,
\end{split}
\ee
where $\partial_1$, $\partial_2$ and $\partial_{12}$ are shorthands for $\partial_{x_1}$, $\partial_{x_2}$ and their difference, and where at the end of the computation one is to set $x_1$ and $x_2$ equal to $x$ and $\zeta_i=0$.
In \cite{st}, where the reader can find more details, we also presented an educated guess for the counterpart of eq.~\eqref{cubicmassless} for the superstring, where \eqref{bosecurents} leaves way to bosonic and fermionic currents. Notice that the spin--2 currents built from a pair of spin--$s$ fields contain at least 2$s$-2 derivatives, and therefore \emph{are not} the energy--momentum tensors of the spin--$s$ fields implied by the Lagrangians of Section \ref{sec:free}.

At this stage we have really left String Theory to return to a field theory framework, albeit in an apparently incomplete fashion, since the starting point was provided by cubic on--shell string amplitudes. Complete cubic vertices that are gauge invariant up to the Fronsdal equations \eqref{ssred} or their compensator analogs \eqref{ssredc} can be recovered completing eq.~\eqref{cubicmassless}, as in \cite{st,mmr}, with two types of terms that we ignored so far. These involve divergences and traces of the three external fields, and actually fall into the combinations
\be
{\cal H}_{ij} \, = \, \left( 1 \, +\, \partial_{\xi_{\,i}}\cdot\partial_{\xi_{\,j}} \right) \, i\, {\cal D}_j \, - \, \frac{1}{2} \, p_j  \cdot \partial_{\xi_{\,i}} \
\partial_{\xi_{\,j}}\cdot\partial_{\xi_{\,j}} \ ,
\ee
with
\be
i\, {\cal D}_i \, = \, p_i \cdot \partial_{\xi_{\,i}} \, - \, \frac{1}{2} \ p_i \cdot \xi_i \ \partial_{\xi_{\,i}} \cdot \partial_{\xi_{\,i}}
\ee
the de Donder operator. In terms of the ${\cal H}_{ij}$ the complete cubic vertex in Fronsdal's constrained form reads  \footnote{Appendix A of \cite{st} explains how to pass to the compensator formulation of Section \ref{sec:freesymmetric}.}
\be
\begin{split}
\cA^{\,\pm} &= e^{\, \pm\, \Gamma} \, \biggl[1 \, + \, \left(\frac{\alpha^{\, \prime}}{2} \right)\ \big({\cal H}_{12} \, {\cal H}_{13}\, +\, {\cal H}_{21}\, {\cal H}_{23}\, +\, {\cal H}_{31}\, {\cal H}_{32} \big)
\pm \left(\frac{\alpha^{\, \prime}}{2} \right)^\frac{3}{2} \, \times \\ & \big( :{\cal H}_{21} \, {\cal H}_{32} \, {\cal H}_{13} :
-  \,: {\cal H}_{12} \, {\cal H}_{31} \, {\cal H}_{23}:  \big)
\biggr] \varphi_{1}\left(p_{1},\xi_{1}\right)\, \varphi_{2}\left(p_{2},\xi_{2}\right)\, \varphi_{3}\left(p_{3},\xi_{3}\right) \ ,
\end{split}
\ee
where the normal ordering moves de Donder operators to the right in products with $\partial_{\xi_{\,i}}$'s. This is the point where the construction of \cite{st} meets the independent work of Manvelyan, Mkrtchyan and Ruhl \cite{mmr}, who obtained the same cubic vertices from field theory considerations. Strictly speaking, however, in \cite{st} we were more precise with the Chan--Paton symmetry \cite{cp}, which makes full use in general of the sign option in eqs.~\eqref{cubicmassless} or \eqref{cubicmassless2}.


\scss{Four--point functions and beyond}\label{sec:beyondcubic}

In \cite{tar} Taronna brought the discussion forward by some stimulating steps that can potentially lead to further progress. He moved from the observation that the operator of eq.~\eqref{Gamma}, whose exponential generates the whole sequence of cubic vertices, comprises two different groups of familiar terms. The first,
\be
(\partial_{\xi_{\,1}}\cdot\partial_{\xi_{\,2}})(\partial_{\xi_{\,3}}\cdot p_{\,12})\,+\,(\partial_{\xi_{\,2}}\cdot\partial_{\xi_{\,3}})
(\partial_{\xi_{\,1}}\cdot p_{\,23})\,+\,(\partial_{\xi_{\,3}}\cdot\partial_{\xi_{\,1}})(\partial_{\xi_{\,2}}\cdot p_{\,31}) \ ,
\ee
in the usual tensor notation would become the familiar cubic Yang--Mills vertex
\be
\left({\rm YM3}\right)_{\mu\nu\rho} \, = \, \eta_{\mu\nu} \, p_{12, \, \rho} \, +\, \eta_{\nu\rho} \, p_{23, \, \mu} \, +\, \eta_{\rho\mu} \, p_{31, \, \nu}  \ , \label{YM3}
\ee
while the rest is a combination of scalar--scalar--vector vertices. Let me stress that the key property of $\Gamma$ is that it commutes with gauge transformations, up to Virasoro or DFP conditions. As we have seen, this makes it possible, for an action combining the quadratic terms of Section \ref{sec:freesymmetric} with the vertex \eqref{cubicmassless}, to be gauge invariant under transformations that are generally deformed. While String Theory uses the exponential of $\Gamma$, it would seem possible to consider more general functions of this operator in HS constructions, altering the relative weights of terms involving different numbers of derivatives. This is another point that deserves further investigation.

A key question addressed in \cite{tar} is what should be the counterparts, for four and higher--point functions, of the exponent $\Gamma$ that builds cubic HS vertices. For brevity, let me leave aside the scalar--scalar--vector terms present in $\Gamma$ to concentrate on the four--point functions built starting solely from \eqref{YM3}. To begin with, these are four--point Yang--Mills S--matrix amplitudes, but interestingly they suffice to define a whole class of HS amplitudes. In order to display them, let us first recast the Yang--Mills amplitudes in terms of \emph{independent} group--theory factors, or
independent combinations of Chan--Paton factors, that a reasoning along the lines of \cite{cp} would guarantee to also allow the factorization of S--matrix amplitudes involving generic HS states. The end result combines exchange amplitudes in pairs of channels and quartic Yang--Mills interactions and thus, amusingly, has \emph{planar duality}, albeit in a very simple form where the poles are manifest in pairs of channels. It is uniquely determined once, making use of the Jacobi identity, one sorts out the contributions to the Yang--Mills amplitude that carry along independent products of structure constants. One of these contributions takes the form
\be
A_\mu(p_1)\, A_\nu(p_2)\, A_\rho(p_3)\, A_\sigma(p_4)\ \cA^{\mu\nu\rho\sigma} \ {\rm tr} \left(\left[ T_1,T_2\right]\left[T_3,T_4\right] \right)  \ , \label{suamplitude}
\ee
where the dynamical amplitude has been combined with its Chan--Paton factors and, schematically,
\be
\cA^{\mu\nu\rho\sigma} \ = \  \left({\rm YM3}\ \frac{1}{s} \ {\rm YM3} \, + \, {\rm YM4}\right)^{\mu\nu\rho\sigma} \ +  \
 \left({\rm YM3}\ \frac{1}{u} \ {\rm YM3} \, + \, {\rm YM4}\right)^{\mu\sigma\nu\rho}  \ . \label{su}
\ee
Here $s$ and $u$ are Mandelstam variables, while the ${\rm YM4}$'s are the usual quartic Yang--Mills couplings. Notice that the decoupling of longitudinal polarizations from \eqref{su} translates into an invariance of \eqref{suamplitude} under {on--shell Abelian gauge transformations} of the type
\be
A_i(p_i) \ \to \ A_i(p_i) \ + \ i\, p_i \, \Lambda(p_i) \ . \label{shifts}
\ee
This key property of Yang--Mills tree amplitudes must hold individually for the two groups of terms, since they carry along independent group theory factors, and in the complete amplitude the $(s,u)$ contribution \eqref{su} is accompanied by a similar $(s,t)$ contribution,
\be
A_\mu(p_1)\, A_\nu(p_2)\, A_\rho(p_3)\, A_\sigma(p_4)\ \cA^{\mu\rho\sigma\nu} \ {\rm tr} \left(\left[ T_1,T_3\right]\left[T_4,T_2\right] \right)  \ . \label{stamplitude}
\ee

The spin--$s$ amplitudes proposed in \cite{tar} include
\be
\varphi_{\mu_1 \ldots \mu_s}(p_1) \, \varphi_{\nu_1 \ldots \nu_s}(p_2)\, \varphi_{\rho_1 \ldots \rho_s}(p_3)\, \varphi_{\sigma_1 \ldots \sigma_s}(p_4) \ \left(su\right)^{s-1} \prod_{k=1}^s \cA^{\mu_k\nu_k\rho_k\sigma_k} \ , \label{4sampl}
\ee
that should be properly dressed with the same Chan--Paton factors as in eq.~\eqref{suamplitude} and then combined with corresponding $(s,t)$ contributions. Notice that these amplitudes are invariant by construction under the HS counterparts of eq.~\eqref{shifts}, Abelian on--shell transformations that in the index--free notation of Section \ref{sec:free} would read
\be
\vf(p_i) \to \vf(p_i) + i\, p_i \, \Lambda(p_i) \ .
\ee
Moreover, the crucial power of $su$ has a two--fold effect: it guarantees that only single poles in $s$ and $u$ be present and raises the spin of the exchanged particles to an \emph{odd} value, $2s-1$, which is higher than the spin of the external states for $s>1$. Notice that the end result possesses the same symmetries as the Chan--Paton factor of eq.~\eqref{suamplitude}, which it should carry in a construction along the lines of \cite{cp}, where the flip symmetries would be directly correlated to the spins of the massless external states.

Tensoring the independent contributions to four--point Yang--Mills amplitudes has thus led to an infinite family of HS amplitudes that by construction do not suffer from Weinberg's factorization problem. How can this be the case? In order to answer this question, Weinberg's argument was extended in \cite{tar} to generic soft amplitudes, taking into account the explicit form of the vertex \eqref{cubicmassless}. The conclusion is that \emph{no problems are encountered if the intermediate spins are not lower than the external ones}, precisely as was the case for eq.~\eqref{4sampl}. The very existence of the amplitudes \eqref{4sampl}, however, raises another important question. Namely, once the cubic vertices are given, what prevents one from building with them ``bad'' amplitudes, with internal spins that are too low to comply with Weinberg's factorization? The answer proposed in \cite{tar} is very interesting and controversial at the same time. It makes use of non--local quartic vertices, in fact the minimal set of them needed to barely remove the ``bad'' exchange amplitudes built from any given pair of vertices \eqref{cubicmassless}. In this fashion, one would be left with a subset of non--vanishing amplitudes, all invariant by construction under the Abelian gauge transformations that signal the decoupling of unphysical polarizations and thus complying, a fortiori, with the extension of Weinberg's soft emission argument. As we have already mentioned, however, in this fashion external HS particles would interact only via exchanges of other HS particles with spins not lower than the external ones! This state of affairs brings to one's mind the peculiar superstring states considered in \cite{dima}, and it would be interesting to elaborate further on the possible links between the two situations. Unfortunately, as noticed in \cite{tar}, with finitely many spins the removal of the unwanted exchanges is apparently in conflict with unitarity, which is somehow the Coleman--Mandula argument haunting back again. A proper understanding of the singular behavior of infinitely many HS fields once all infrared cutoffs, be they masses or a cosmological term, are removed, will thus require more effort. Clearly, further progress along these lines will be highly instructive.




\vskip 12pt


\section*{Acknowledgments}


This work was supported in part by Scuola Normale Superiore, by INFN, by the MIUR-PRIN contract 2009-KHZKRX and by the ERC Senior Grant n. 226455, "Supersymmetry, Quantum Gravity and Gauge Fields" (SUPERFIELDS). I am very grateful to Mirian Tsulaia and Jihad Mourad for pleasant and instructive collaborations on some of the topics reviewed in these notes, and to Dario Francia, Andrea Campoleoni and Massimo Taronna for long and fruitful collaborations and also for a number of useful suggestions on the manuscript, whose final form also benefitted from some input from Karapet Mkrtchyan. Finally, I would like to thank Prof.~Noriaki Kitazawa of Tokyo Metropolitan University for a careful ``non--expert reading'' that helped me greatly to streamline the presentation.








\vskip 24pt

\begin{thebibliography}{99}

\bibitem{veneziano}
G.~Veneziano,
  Nuovo Cim.\  {\bf A57 } (1968)  190-197.

\bibitem{stringtheory}
See, for instance:
M.~B.~Green, J.~H.~Schwarz and E.~Witten, ``Superstring Theory'', 2
vols. (Cambridge Univ. Press, Cambridge, UK, 1987); J.~Polchinski,
``String theory'', 2 vols. (Cambridge Univ. Press, Cambridge, UK,
1998); B.~Zwiebach, ``A first course in string theory'', (Cambridge
Univ. Press, Cambridge, UK, 2004); K.~Becker, M.~Becker and J.~H.~Schwarz, ``String theory and
M-theory: A modern introduction'', (Cambridge Univ. Press, Cambridge, UK, 2007); E.~Kiritsis,
``String theory in a nutshell'', (Princeton Univ. Press, Princeton, NJ, USA, 2007).

\bibitem{solvay}
``Higher-Spin Gauge Theories'', Proceedings of the First Solvay Workshop, held in
Brussels on May 12-14, 2004, eds. R.~Argurio, G.~Barnich, G.~Bonelli and M.~Grigoriev
(Int. Solvay Institutes, 2006). This collection contains some contributions closely
related, in spirit, to the present work, including: M.~Bianchi and V.~Didenko,
arXiv:hep-th/0502220;
D.~Francia and C.~M.~Hull,
arXiv:hep-th/0501236;
N.~Bouatta, G.~Compere and A.~Sagnotti,
arXiv:hep-th/0409068;
X.~Bekaert, S.~Cnockaert, C.~Iazeolla and M.~A.~Vasiliev,
arXiv:hep-th/0503128;
A.~Sagnotti, E.~Sezgin and P.~Sundell,
arXiv:hep-th/0501156.
Other recent reviews include:
D.~Sorokin,
  AIP Conf.\ Proc.\ {\bf 767} (2005) 172
  [hep-th/0405069];
D.~Francia and A.~Sagnotti,
J.\ Phys.\ Conf.\ Ser.\  {\bf 33} (2006) 57
[arXiv:hep-th/0601199];
A.~Fotopoulos and M.~Tsulaia,
arXiv:hep-th/0805.1346;
  A.~Campoleoni,
  Riv.\ Nuovo Cim.\  {\bf 033 } (2010)  123-253.
  [arXiv:0910.3155 [hep-th]];
A.~Sagnotti,
arXiv:hep-th/1002.3388;
  X.~Bekaert, N.~Boulanger and P.~Sundell,
  arXiv:1007.0435 [hep-th].

\bibitem{sftheory}
W.~Siegel,
Nucl.\ Phys.\ {\bf B263} (1986) 93;
W.~Siegel and B.~Zwiebach,
Nucl.\ Phys.\ {\bf B263} (1986) 105;
T.~Banks and M.~E.~Peskin,
Nucl.\ Phys.\ {\bf B264} (1986) 513;
M.~Kato and K.~Ogawa,
Nucl.\ Phys.\ {\bf B212} (1983) 443;
N.~Ohta,
Phys.\ Rev.\ {\bf D33} (1986) 1681,
Phys.\ Lett.\ {\bf B179} (1986) 347,
Phys.\ Rev.\ Lett.\  {\bf 56} (1986) 440
[Erratum-ibid.\  {\bf 56} (1986) 1316];
A.~Neveu, H.~Nicolai and P.~C.~West,
Nucl.\ Phys.\ {\bf B264} (1986) 573;
A.~Neveu and P.~C.~West,
Nucl.\ Phys.\ {\bf B268} (1986) 125; E.~Witten,
Nucl.\ Phys.\ {\bf B268} (1986) 253.

\bibitem{bbms}
M.~Bianchi, J.~F.~Morales and H.~Samtleben,
  JHEP\ {\bf 0307} (2003) 062
  [arXiv:hep-th/0305052 [hep-th]];
 N.~Beisert, M.~Bianchi, J.~F.~Morales and H.~Samtleben,
  JHEP\ {\bf 0402} (2004) 001
  [hep-th/0310292],
  JHEP\ {\bf 0407} (2004) 058
  [hep-th/0405057].

\bibitem{adscft}
  J.~M.~Maldacena,
  Adv.\ Theor.\ Math.\ Phys.\  {\bf 2 } (1998)  231-252.
  [hep-th/9711200];
    E.~Witten,
  Adv.\ Theor.\ Math.\ Phys.\  {\bf 2 } (1998)  253-291.
  [hep-th/9802150];
    S.~S.~Gubser, I.~R.~Klebanov, A.~M.~Polyakov,
  Phys.\ Lett.\  {\bf B428 } (1998)  105-114.
  [hep-th/9802109].
  For a review see:
    O.~Aharony, S.~S.~Gubser, J.~M.~Maldacena, H.~Ooguri, Y.~Oz,
  Phys.\ Rept.\  {\bf 323 } (2000)  183-386.
  [hep-th/9905111].

\bibitem{majorana}
E.~Majorana,
Nuovo Cim.\  {\bf 9} (1932) 335.

\bibitem{dfp}
P.~A.~M.~Dirac,
Proc.\ Roy.\ Soc.\ Lond.\  {\bf 155A} (1936) 447;
M.~Fierz,
  Helv.\ Phys.\ Acta {\bf 12 } (1939)  3-37;
M.~Fierz and W.~Pauli,
Proc.\ Roy.\ Soc.\ Lond.\  {\bf A173} (1939) 211.

\bibitem{vasiliev}
M.~A.~Vasiliev,
Annals Phys.\  {\bf 190} (1989) 59,
Phys.\ Lett.\  {\bf B238} (1990) 305,
Phys.\ Lett.\  {\bf B243} (1990) 378,
Class.\ Quant.\ Grav.\  {\bf 8} (1991) 1387,
Phys.\ Lett.\  {\bf B257} (1991) 111,
Phys.\ Lett.\  {\bf B285} (1992) 225,
Fortsch.\ Phys.\  {\bf 52} (2004) 702
[arXiv:hep-th/0401177];
\\
For a review see also:
M.~A.~Vasiliev,
Int.\ J.\ Mod.\ Phys.\  {\bf D5} (1996) 763
[arXiv:hep-th/9611024],
arXiv:hep-th/9910096.


\bibitem{fms}
D.~Francia, J.~Mourad and A.~Sagnotti,
Nucl.\ Phys.\  {\bf B773} (2007) 203
[arXiv:hep-th/0701163];
  Nucl.\ Phys.\  {\bf B804 } (2008)  383-420.
  [arXiv:0803.3832 [hep-th]].

\bibitem{old}
T.~Curtright,
  Phys.\ Lett.\  {\bf B165} (1985) 304;
  C.~S.~Aulakh, I.~G.~Koh and S.~Ouvry,
  Phys.\ Lett.\  {\bf B173} (1986) 284;
C.~S.~Aulakh, I.~G.~Koh and S.~Ouvry,
  Phys.\ Lett.\  {\bf B173} (1986) 284;
S.~Ouvry and J.~Stern,
  Phys.\ Lett.\  {\bf B177} (1986) 335;
  W.~Siegel and B.~Zwiebach,
  Nucl.\ Phys.\  {\bf B282} (1987) 125;
W.~Siegel,
  Nucl.\ Phys.\  {\bf B284} (1987) 632;

\bibitem{labastida}
J.~M.~F.~Labastida and T.~R.~Morris,
Phys.\ Lett.\  {\bf B180} (1986) 101;
J.~M.~F.~Labastida,
Phys.\ Rev.\ Lett.\  {\bf 58} (1987) 531;
J.~M.~F.~Labastida,
Nucl.\ Phys.\  {\bf B322} (1989) 185.

\bibitem{nlocmixed}
X.~Bekaert and N.~Boulanger,
Commun.\ Math.\ Phys.\  {\bf 245} (2004) 27
[arXiv:hep-th/0208058],
Phys.\ Lett.\  {\bf B561} (2003) 183
[arXiv:hep-th/0301243];
P.~de Medeiros and C.~Hull,
JHEP {\bf 0305} (2003) 019
[arXiv:hep-th/0303036].

\bibitem{mixed}
A.~Campoleoni, D.~Francia, J.~Mourad and A.~Sagnotti,
Nucl.\ Phys.\  {\bf B815} (2009) 289
[arXiv:hep-th/0810.4350],
Nucl.\ Phys.\  {\bf B828} (2010) 405
[arXiv:hep-th/0904.4447].

\bibitem{ff}
C.~Fronsdal,
Phys.\ Rev.\  {\bf D18} (1978) 3624;
J.~Fang and C.~Fronsdal,
Phys.\ Rev.\  {\bf D18} (1978) 3630.

\bibitem{fs1}
D.~Francia and A.~Sagnotti,
Phys.\ Lett.\  {\bf B543} (2002) 303
[arXiv:hep-th/0207002],
Class.\ Quant.\ Grav.\  {\bf 20} (2003) S473
[arXiv:hep-th/0212185];
Phys.\ Lett.\  {\bf B624} (2005) 93
[arXiv:hep-th/0507144].

\bibitem{triplet}
A.~K.~H.~Bengtsson,
Phys.\ Lett.\  {\bf B182} (1986) 321;
M.~Henneaux and C.~Teitelboim, in ``Quantum Mechanics of Fundamental Systems, 2'',
eds. C. Teitelboim and J. Zanelli (Plenum Press, New York, 1988), p. 113;
G.~Bonelli,
Nucl.\ Phys.\  {\bf B669 } (2003)  159-172.
[hep-th/0305155];
A.~K.~H.~Bengtsson,
Class.\ Quant.\ Grav.\  {\bf 5} (1988)  437;
A.~Fotopoulos and M.~Tsulaia,
Int.\ J.\ Mod.\ Phys.\  {\bf A24} (2009) 1
[arXiv:0805.1346 [hep-th]].

\bibitem{wigner}
E.~P.~Wigner,
  Annals Math.\  {\bf 40 } (1939)  149-204.

\bibitem{regge}
  T.~Regge,
  Nuovo Cim.\  {\bf 14 } (1959)  951.

\bibitem{gunaydin}
For a review see:
  M.~Gunaydin,
  Lect.\ Notes Phys.\  {\bf 180 } (1983)  192-213.

\bibitem{diracsing}
  P.~A.~M.~Dirac,
  J.\ Math.\ Phys.\  {\bf 4 } (1963)  901-909.

\bibitem{flatofronsdal}
  C.~Fronsdal,
  Rev.\ Mod.\ Phys.\  {\bf 37 } (1965)  221-224;
  M.~Flato, C.~Fronsdal,
  Lett.\ Math.\ Phys.\  {\bf 2 } (1978)  421-426;
    E.~Angelopoulos, M.~Flato, C.~Fronsdal, D.~Sternheimer,
  Phys.\ Rev.\  {\bf D23 } (1981)  1278.

\bibitem{casalbuoni}
  R.~Casalbuoni,
  PoS {\bf EMC2006 } (2006)  004.
  [arXiv:hep-th/0610252 [hep-th]];
    X.~Bekaert, M.~R.~de Traubenberg, M.~Valenzuela,
  JHEP {\bf 0905 } (2009)  118.
  [arXiv:0904.2533 [hep-th]];
   S.~Esposito,
  [arXiv:1110.6878 [physics.hist-ph]].

\bibitem{RS}  W.~Rarita, J.~Schwinger,
  Phys.\ Rev.\  {\bf 60 } (1941)  61.

\bibitem{supergravity}
  D.~Z.~Freedman, P.~van Nieuwenhuizen and S.~Ferrara,
  Phys.\ Rev.\ {\bf D13} (1976) 3214;
    S.~Deser and B.~Zumino,
  Phys.\ Lett.\ {\bf B62} (1976) 335.

\bibitem{weinb64}
  S.~Weinberg,
  Phys.\ Rev.\  {\bf 135 } (1964)  B1049-B1056.

\bibitem{cm}   S.~R.~Coleman, J.~Mandula,
  Phys.\ Rev.\  {\bf 159 } (1967)  1251-1256.

\bibitem{vz}
  G.~Velo and D.~Zwanziger,
  Phys.\ Rev.\  {\bf 186} (1969) 1337,
  Phys.\ Rev.\  {\bf 188} (1969) 2218;
  G.~Velo,
  Nucl.\ Phys.\  {\bf B43} (1972) 389.

\bibitem{ad}
  C.~Aragone and S.~Deser,
Phys.\ Lett.\  {\bf B86} (1979) 161,
Nuovo Cim.\  {\bf B57} (1980) 33.

\bibitem{ww}
S.~Weinberg, E.~Witten,
  Phys.\ Lett.\  {\bf B96 } (1980)  5.

\bibitem{porratiww}
  M.~Porrati,
  Phys.\ Rev.\  {\bf D78 } (2008)  065016
  [arXiv:0804.4672 [hep-th]].

\bibitem{SH}
L.~P.~S.~Singh and C.~R.~Hagen,
Phys.\ Rev.\  {\bf D9} (1974) 910,
Phys.\ Rev.\  {\bf D9} (1974) 898.

\bibitem{dubna}
A.~Pashnev and M.~M.~Tsulaia,
Mod.\ Phys.\ Lett.\  {\bf A12} (1997) 861
[arXiv:hep-th/9703010];
Mod.\ Phys.\ Lett.\  {\bf A13} (1998) 1853
[arXiv:hep-th/9803207],
C.~Burdik, A.~Pashnev and M.~Tsulaia,
Nucl.\ Phys.\ Proc.\ Suppl.\  {\bf 102} (2001) 285
[arXiv:hep-th/0103143];
I.~L.~Buchbinder, A.~Pashnev and M.~Tsulaia,
Phys.\ Lett.\  {\bf B523} (2001) 338
[arXiv:hep-th/0109067]
arXiv:hep-th/0206026;
X.~Bekaert, I.~L.~Buchbinder, A.~Pashnev and M.~Tsulaia,
Class.\ Quant.\ Grav.\  {\bf 21} (2004) S1457
[arXiv:hep-th/0312252];
I.~L.~Buchbinder, V.~A.~Krykhtin and A.~Pashnev,
Nucl.\ Phys.\  {\bf B711} (2005) 367
[arXiv:hep-th/0410215].

\bibitem{brst}
C.~Becchi, A.~Rouet and R.~Stora,
Commun.\ Math.\ Phys.\  {\bf 42} (1975) 127,
Annals Phys.\  {\bf 98} (1976) 287;
I.V. Tyutin, preprint FIAN n. 39 (1975);
E.~S.~Fradkin and G.~A.~Vilkovisky,
Phys.\ Lett.\ {\bf B55} (1975) 224;
I.~A.~Batalin and G.~A.~Vilkovisky,
Phys.\ Lett.\ {\bf B69} (1977) 309;
M.~Henneaux,
Phys.\ Rept.\  {\bf 126} (1985) 1.

\bibitem{schwinger}
J.~S.~Schwinger,
  Reading, USA: Addison-Wesley (1989) 306 p. (Advanced book classics series).

\bibitem{lowder}
D.~Francia,
  Nucl.\ Phys.\ {\bf B796 } (2008)  77-122.
  [arXiv:0710.5378 [hep-th]],
  Fortsch.\ Phys.\  {\bf 56 } (2008)  800-808.
  [arXiv:0804.2857 [hep-th]].

\bibitem{bgk}
I.~L.~Buchbinder, A.~V.~Galajinsky, V.~A.~Krykhtin,
  Nucl.\ Phys.\  {\bf B779 } (2007)  155-177.
  [hep-th/0702161];
  I.~L.~Buchbinder, V.~A.~Krykhtin, A.~A.~Reshetnyak,
  Nucl.\ Phys.\  {\bf B787 } (2007)  211-240.
  [hep-th/0703049];
  I.~L.~Buchbinder, A.~V.~Galajinsky,
  JHEP {\bf 0811 } (2008)  081.
  [arXiv:0810.2852 [hep-th]].

\bibitem{dwf}
B.~de Wit and D.~Z.~Freedman,
Phys.\ Rev.\  {\bf D21} (1980) 358.

\bibitem{st1}
A.~Sagnotti and M.~Tsulaia,
Nucl.\ Phys.\  {\bf B682} (2004) 83 [arXiv:hep-th/0311257].

\bibitem{francia10}
D.~Francia,
  J.\ Phys.\ Conf.\ Ser.\  {\bf 222 } (2010)  012002.
  [arXiv:1001.3854 [hep-th]];
Phys.\ Lett.\  {\bf B690} (2010) 90
[arXiv:1001.5003 [hep-th]].

\bibitem{tar}
M.~Taronna, ``Higher-Spin Interactions:~four-point functions and beyond,''
[arXiv:1107.5843 [hep-th]].

\bibitem{st}
M.~Taronna,
``Higher Spins and String Interactions'', Master Thesis,
arXiv:1005.3061 [hep-th];
A.~Sagnotti, M.~Taronna,
Nucl.\ Phys.\  {\bf B842 } (2011)  299-361.
[arXiv:1006.5242 [hep-th]].

\bibitem{mmr}
R.~Manvelyan, K.~Mkrtchyan, W.~Ruhl,
Nucl.\ Phys.\  {\bf B836 } (2010)  204-221.
[arXiv:1003.2877 [hep-th]];
  Phys.\ Lett.\ {\bf B696} (2011) 410
  [arXiv:1009.1054 [hep-th]];
R.~Manvelyan, K.~Mkrtchyan, W.~Ruhl, M.~Tovmasyan,
Phys.\ Lett.\  {\bf B699 } (2011)  187-191.
[arXiv:1102.0306 [hep-th]].
For a review, see:   K.~Mkrtchyan,
  arXiv:1101.5643 [hep-th].


\bibitem{argnap}
  P.~C.~Argyres, C.~R.~Nappi,
  Phys.\ Lett.\  {\bf B224 } (1989)  89.

\bibitem{prs}
M.~Porrati, R.~Rahman, A.~Sagnotti,
  Nucl.\ Phys.\  {\bf B846 } (2011)  250-282.
  [arXiv:1011.6411 [hep-th]].

\bibitem{Fradkin}
E.~S.~Fradkin and M.~A.~Vasiliev,
Nucl.\ Phys.\ {\bf B291} (1987) 141,
Phys.\ Lett.\ {\bf B189} (1987) 89.

\bibitem{cp}
J.~E.~Paton and H.~M.~Chan,
Nucl.\ Phys.\  {\bf B10} (1969) 516;
J.~H.~Schwarz,
``Gauge Groups For Type I Superstrings'', CALT-68-906-REV
\textit{Presented at 6th Johns Hopkins Workshop on Current Problems in High-Energy Particle Theory, Florence, Italy, Jun 2-4, 1982};
N.~Marcus and A.~Sagnotti,
Phys.\ Lett.\  {\bf B119} (1982) 97,
Phys.\ Lett.\  {\bf B188} (1987) 58.
For reviews see:
 J.~H.~Schwarz,
  Phys.\ Rept.\  {\bf 89 } (1982)  223-322;
C.~Angelantonj, A.~Sagnotti,
  Phys.\ Rept.\  {\bf 371 } (2002)  1-150.
  [hep-th/0204089].

\bibitem{metsaev}
R.~R.~Metsaev,
arXiv:hep-th/0712.3526,
Nucl.\ Phys.\  {\bf B759} (2006) 147
[arXiv:hep-th/0512342].

\bibitem{goteborg}
  A.~K.~H.~Bengtsson, I.~Bengtsson and L.~Brink,
  Nucl.\ Phys.\ {\bf B227} (1983) 31, 41.
\bibitem{bls}
N.~Boulanger, S.~Leclercq and P.~Sundell,
JHEP {\bf 0808} (2008) 056
[arXiv:hep-th/0805.2764].

\bibitem{rome09}
A.~Sagnotti, talk at Strings 09, http://strings2009.roma2.infn.it/program.html.

\bibitem{vdvz}
  H.~van Dam and M.~J.~G.~Veltman,
  Nucl.\ Phys.\ {\bf B22}, 397 (1970);
  V.~I.~Zakharov,
  JETP Lett.\  {\bf 12} (1970) 312
  [Pisma Zh.\ Eksp.\ Teor.\ Fiz.\  {\bf 12} (1970) 447].

\bibitem{bjm}
X.~Bekaert, E.~Joung and J.~Mourad,
JHEP {\bf 0905} (2009) 126
[arXiv:hep-th/0903.3338].
JHEP {\bf 1102} (2011) 048
[arXiv:1012.2103 [hep-th]].

\bibitem{hp}
  A.~Higuchi,
  Nucl.\ Phys.\ {\bf B282} (1987) 397,
  J.\ Math.\ Phys.\  {\bf 28} (1987) 1553
  [Erratum-ibid.\  {\bf 43} (2002) 6385];
  M.~Porrati,
  Phys.\ Lett.\  {\bf B498}, 92 (2001)
  [arXiv:hep-th/0011152];
I.~I.~Kogan, S.~Mouslopoulos and A.~Papazoglou,
  Phys.\ Lett.\ {\bf B503}, 173 (2001)
  [arXiv:hep-th/0011138].

\bibitem{shortening}
J. Dixmier,
``Repr\'esentations int\'egrables du groupe de De Sitter'', Bulletin
de la Soci\'et\'e Math\'ematique de France, 89 (1961), p. 9-41.
Earlier works on the representations of SO(1,4) include: L.~Thomas,
    ``On unitary representations of the group of the De Sitter group,"
    Ann.\ of Math.\ {\bf 42} (1941) 113;
    T. Newton,
    ``A note on the representation of the De Sitter group,"
    Ann.\ of  Math.\ {\bf 52} (1950) 730;
A.~W.~Knapp and E.~M.~Stein,
    ``Interwining operators for semisimple groups,"
    Ann.\ of Math.\ (2) {\bf 93} (1971) 489;
    E.~Thieleker,
    ``The unitary representations of the generalized Lorentz groups,"
    Transactions of the American mathematical Society {\bf 199} (1974)
    327;
    N.~Ja.~Vilenkin and A.~U.~Kilmyk,
    ``Representation of Lie groups and special fucntions,"
    Kluwer Academic Publishers (1993). The two-dimensional case was
    considered in
  V.~Bargmann,
  Annals Math.\  {\bf 48} (1947) 568.

  \bibitem{pgauges}
S.~Deser and R.~I.~Nepomechie,
  Annals Phys.\  {\bf 154} (1984) 396;
  S.~Deser and A.~Waldron,
  Phys.\ Rev.\ Lett.\  {\bf 87} (2001) 031601
  [arXiv:hep-th/0102166],
  Nucl.\ Phys.\ {\bf B607} (2001) 577
  [arXiv:hep-th/0103198],
  Phys.\ Lett.\ {\bf B501}, 134 (2001)
  [arXiv:hep-th/0012014];

\bibitem{radial}
C.~Fronsdal,
  Phys.\ Rev.\ {\bf D20} (1979) 848.
  J.~Fang and C.~Fronsdal,
  Phys.\ Rev.\ {\bf D22} (1980) 1361;
  T.~Biswas and W.~Siegel,
  JHEP {\bf 0207} (2002) 005
  [arXiv:hep-th/0203115];
  K.~Hallowell and A.~Waldron,
  Nucl.\ Phys.\ {\bf B724} (2005) 453
  [arXiv:hep-th/0505255];
X.~Bekaert, I.~L.~Buchbinder, A.~Pashnev and M.~Tsulaia,
  Class.\ Quant.\ Grav.\  {\bf 21} (2004) S1457
  [arXiv:hep-th/0312252].

\bibitem{whiwat}
See, for instance, E.S. Whittaker and G.N. Watson, \emph{A Course of Modern Analysis} (Cambridge Univ. Press, Cambridge, 1973).

\bibitem{bbvd}
F.~A.~Berends, G.~J.~H.~Burgers and H.~van Dam,
Nucl.\ Phys.\ {\bf B271} (1986) 429,
Z.\ Phys.\ {\bf C24} (1984) 247,
Nucl.\ Phys.\ {\bf B260} (1985) 295.

\bibitem{grossmende}
D.~J.~Gross and P.~F.~Mende,
Phys.\ Lett.\ {\bf B197} (1987) 129,
Nucl.\ Phys.\ {\bf B303} (1988) 407.

\bibitem{adsembedding}
  E.~Joung and M.~Taronna,
  arXiv:1110.5918 [hep-th].

\bibitem{dima}
  D.~Polyakov,
  Phys.\ Rev.\ {\bf D82} (2010) 066005
  [arXiv:0910.5338 [hep-th]];
  Int.\ J.\ Mod.\ Phys.\ {\bf A25} (2010) 4623
  [arXiv:1005.5512 [hep-th]].

\end{thebibliography}
\end{document}